\documentclass[reprint,amsmath,amssymb,aps,nofootinbib, superscriptaddress]{revtex4-2}

\def\bel#1{\begin{equation} \label{#1}}

\def\mpl{M_{\rm pl}}

\def\c{\chi_b}

\def\be{\begin{equation}}
\def\ee{\end{equation}}
\def\bea{\begin{eqnarray}}
\def\eea{\end{eqnarray}}

\def\ltap{\ \raise.3ex\hbox{$<$\kern-.75em\lower1ex\hbox{$\sim$}}\ }
\def\gtap{\ \raise.3ex\hbox{$>$\kern-.75em\lower1ex\hbox{$\sim$}}\ }
\def\gl{\ \raise.5ex\hbox{$>$}\kern-.8em\lower.5ex\hbox{$<$}\ }
\def\roughly#1{\raise.3ex\hbox{$#1$\kern-.75em\lower1ex\hbox{$\sim$}}}

\def\pref#1{(\ref{#1})}

\def\mpl{M_{\rm pl}}

\def\varp{\varphi}

\def\mv{m_{\varphi}}

\newcommand{\comments}[1]{}
\newcommand{\nef}{\Delta N_{\rm eff}}

\newcommand{\mef}{m_{\rm sp}^{\rm eff}}
\newcommand{\s}{\hat{s}}
\newcommand{\ben}{\begin{enumerate}}
\newcommand{\een}{\end{enumerate}}
\newcommand{\bi}{\begin{itemize}}
\newcommand{\ei}{\end{itemize}}
\newcommand{\ba}{\begin{align}}
\newcommand{\ea}{\end{align}}

\def\x{{\bf \textit{x}}}

\def\beq{\begin{equation}}
\def\eeq{\end{equation}}
\def\bea{\begin{eqnarray}}
\def\eea{\end{eqnarray}}

\usepackage[dvips]{graphicx} %
\usepackage{graphicx,amsmath,amsfonts,amssymb,slashed,float,hyperref}
\usepackage[normalem]{ulem}
\usepackage{bbold,wasysym}
\usepackage{graphicx}
\usepackage{aas_macros}
\usepackage{array,multirow}
\usepackage[utf8]{inputenc}

\usepackage[usenames,dvipsnames]{xcolor} 

\usepackage{soul}

\begin{document}

\title{Non-thermal neutrino-like hot dark matter in light of the $S_8$ tension}

\author{Subinoy Das}
\email{subinoy@iiap.res.in}
\affiliation{Indian Institute of Astrophysics, Bengaluru, Karnataka 560034, India}

\author{Anshuman Maharana}
\email{anshumanmaharana@hri.res.in}
\affiliation{Harish-Chandra Research Institute, HBNI, Allahabad 211019, India.}

\author{Vivian Poulin}
\email{vivian.poulin@umontpellier.fr}
\affiliation{Laboratoire Univers \& Particules de Montpellier (LUPM),
CNRS \& Universit\'{e} de Montpellier (UMR-5299),
Place Eug\`{e}ne Bataillon, F-34095 Montpellier Cedex 05, France}

\author{Ravi Kumar Sharma}
\email{ravi.sharma@iiap.res.in}
\affiliation{Indian Institute of Astrophysics, Bengaluru, Karnataka 560034, India}

\begin{abstract}
The $\Lambda$CDM prediction of $S_8\equiv\sigma_8(\Omega_m/0.3)^{0.5}$ -- where $\sigma_8$ is the root mean square of matter fluctuations on  8 $h^{-1}$Mpc scale -- once calibrated on Planck CMB data is $2-3\sigma$ lower than its direct estimate by a number of weak lensing surveys. 
In this paper, we explore the possibility that the `$S_8$-tension' is due to a fractional contribution of non-thermal hot dark matter (HDM)  to the energy density of the universe leading to a power suppression at small-scales in the matter power spectrum. Any HDM model can be characterized by  its effective mass $\mef$ and its contribution to the relativistic degrees of freedom at CMB decoupling $\nef$. Taking the specific example of a sterile particle produced from the decay of the inflaton during an early  matter-dominated era, we find that  the tension can be reduced below $2\sigma$ from  Planck data only, but it does not favor a non-zero $\{\mef,\nef\}$. In combination with a measurement of $S_8$ from KIDS1000+BOSS+2dfLenS, the $S_8$-tension would hint at the existence of a particle of mass $\mef\simeq 0.67_{-0.48}^{+0.26}$ ${\rm eV}$  with a contribution to $\nef\simeq0.06\pm0.05$. However, Pantheon and BOSS BAO/$f\sigma_8$ data restricts the particle mass to $\mef\simeq 0.48_{-0.36}^{+0.17}$ and contribution to $\nef \simeq 0.046_{-0.031}^{+0.004}$. We discuss implications of our results for other canonical non-thermal HDM models-- the Dodelson-Widrow model and a hidden sector model of a thermal sterile particle with a different temperature. We report competitive results on such  hidden sector temperature which might have interesting implications for particle physics model building, in particular connecting the $S_8$-tension to the longstanding short baseline oscillation anomaly.
\end{abstract}
\date{\today}

\maketitle

\section{Introduction}

The $\Lambda$ Cold Dark Matter ($\Lambda$CDM) model of cosmology is compelling at describing a wide variety of observations up to a high degree of accuracy despite the  nature of its dominant components -  Cold Dark Matter (CDM) and Dark Energy (DE) -- still being unknown. Nevertheless, in recent years, a number of intriguing discrepancies have emerged between the values of some cosmological parameter predicted within $\Lambda$CDM -- once the model is calibrated onto Planck Cosmic Microwave Background (CMB) data, Baryon Acoustic Oscillation (BAO) and luminosity distance to SuperNovae of type Ia (SNIa) -- and their direct measurements.

At the heart of this study is the longstanding tension affecting the determination of the amplitude of matter fluctuations, typically parameterized as $S_8\equiv\sigma_8(\Omega_m/0.3)^{0.5}$, where $\sigma_8$ is the root mean square of matter fluctuations on an 8 $h^{-1}$Mpc scale,
and $\Omega_m$ is the total matter abundance. The latest prediction from Planck CMB data within the $\Lambda$CDM framework is $S_8=0.832 \pm0.013$ \cite{Cosmo_collaboration2018planck}.
Originally, observations of galaxies through weak lensing by the CFHTLenS collaboration have indicated that the $\Lambda$CDM model predicts a $S_8$ value that is larger than the direct measurement at the $2\sigma$ level \cite{Heymans:2013fya,MacCrann:2014wfa}.  
This tension has since then been further established within the KiDS/Viking data \cite{Hildebrandt:2018yau,Joudaki:2019pmv}, but is milder within the DES data \cite{Abbott:2017wau}. 
However, a re-analysis of the DES data, combined with KiDS/Viking, led to a determination of $S_8$ that is discrepant with Planck at the $3\sigma$ level, $S_8=0.755_{-0.021}^{+0.019}$  \cite{Joudaki:2019pmv}. 
Recently, the combination of KiDS/Viking and SDSS data has established  $S_8=0.766^{+0.02}_{-0.014}$ \cite{Heymans:2020gsg}. Moreover, it is now understood that the tension is driven by a lower matter clustering amplitude $\sigma_8$. This is mainly due to the fact that $\Omega_M$is strongly constrained –even in extension from $\Lambda$CDM– from the
observations of uncalibrated luminosity distance to supernovae and baryonic acoustic oscillations.
This is particularly interesting for model building: resolving the $S_8$ tension requires to decrease the amplitude of matter fluctuations on scales $k\sim 0.1-1 ~h$/Mpc, which can be easily achieved in a variety of models often related to new DM properties \cite{Kumar:2016zpg,Murgia:2016ccp,Archidiacono:2019wdp,DiValentino:2020vvd,Becker:2020hzj,Enqvist:2015ara,Poulin:2016nat,vattis_late_2019,Haridasu:2020xaa,Clark:2020miy,Pandey:2019plg,Abellan:2020pmw,Abellan:2021bpx}, or new neutrino properties \cite{Poulin:2018zxs,Kreisch:2019yzn}.

In this paper, we explore the possibility that the `$S_8$-tension' is due to the existence of a non-thermal  hot dark matter (HDM) consisting of light sterile neutrinos or  hidden sector particles contributing to a fraction of the dark matter (DM) density in the universe, and leading to a power suppression at small-scales in the matter power spectrum. It is well known that just adding a thermal neutrino-like radiation  $\nef$, together with a non-zero neutrino mass $m_\nu$,  does not resolve the $S_8$-problem \cite{Poulin:2018zxs,Cosmo_collaboration2018planck}.
Here, we explore the consequences of a non-thermal momentum distribution for the hot component (or a temperature different from our visible sector), for the $S_8$-tension. 
In practice, we consider the momentum distribution associated
with sterile particles produced from decays during an early matter domination to radiation domination transition of the universe. We refer to the model as $\nu_{\rm NT}\Lambda$CDM.
From the point of view of theoretical models, it is natural for the early universe to enter an epoch of early matter-dominated era (EMDE)  \cite{Kane:2015jia,Kofman:1997yn,Allahverdi:2010xz}. This EMDE epoch transitions to radiation-dominated era through the decay of the inflaton or cold moduli which dominates the energy density of the universe at early times or EMDE can also appear from hidden sector physics \cite{Berlin:2016gtr, Tenkanen:2016jic}.  In string and
theories of supergravity, this occurs due to moduli vacuum misalignment \cite{Coughlan:1983ci, Banks:1993en, deCarlos:1993wie}. For detailed arguments on the generality of this and computations in explicit settings see  e.g Refs.~\cite{randall, koushik, bobby}.

It was shown that the decay products obtain a characteristic
momentum distribution \cite{Miller:2019pss,oturner,jj,decayy}, that is associated with decays
taking place in a matter-dominated universe evolving to radiation domination. 
The momentum distribution function is essentially fixed by the kinematics. Of
course, this happens only under certain conditions, these are as follows: the particles arise from 1 $\rightarrow$ 2 decay of
the unstable particle (whose quanta dominate the energy density of the universe) and have a mass much smaller than the
mass of the decaying particle. Furthermore, the particles will be taken to be inert, so that they free stream after production.
Thermalization of the decay products leads
to the loss of all information about the kinematics of the decay process. But in a setting with a large number of
hidden sectors, one can expect that some of the species produced during the decay do not thermalize due to very weak interactions. Our scenario belongs to a category where a moduli or inflaton field decays to non-thermal sterile particles. There might be other  particles such as Feebly interacting massive particle (FIMP) which can also produce non-thermal or partially thermal neutrino-like particles \cite{Boyarsky:2021yoh}. The presence of non-thermal dark radiation can affect the CMB \cite{Hou:2011ec} as well as large-scale structures in specific ways and can be probed by precision cosmological data.  
 The study of the implications of
sterile particles with this momentum distribution  for precision cosmology was recently initiated\footnote{For earlier in work on
inert particles from decays see  e.g \cite{oturner, jj, cm,Miller:2019pss}} in Ref.~\cite{decayy}.
Given that the effect of massive sterile particles on the CMB and matter power spectra is well-known (e.g. \cite{acero,Lesgourgues:2006nd,Lesgourgues:2011re} for reviews), it was anticipated there one might get a substantial power suppression in the matter power spectrum due to the momentum distribution of the non-thermal decay products. This power suppression has implications for the $S_8$-tension.

  In this article, we perform a comprehensive monte-carlo markov chains (MCMC) analysis against up-to-date data from Planck, BOSS (BAO and redshift space distortions $f\sigma_8$) and Pantheon data, with and without the inclusion of a prior on the value of $S_8$ as measured with the KiDS/Viking+BOSS+2dFLens data\footnote{For analysis in similar spirit (although without inclusion
  of the $S_{8}$) prior motivated by short base line neutrino experiments see  e.g \cite{sr1,sr}. Of course, here
  the momentum distribution of the sterile particles is assumed to be as motivated by neutrino physics i.e thermal or the Dodelson-Widrow distribution \cite{Dodelson:1993je}.}.
We find that the $\nu_{\rm NT}\Lambda$CDM model can indeed alleviate the tension between Planck and $S_8$ measurements, but the success of the resolution is slightly degraded once  BOSS and Pantheon data are included in the analysis.
To better understand the features of the model  leading to a resolution of the tension, we compare the non-thermal sterile neutrino model to the standard massive neutrino model with extra relativistic degrees of freedom.
We find that, for similar effect on the CMB power spectrum, the  $\nu_{\rm NT}\Lambda$CDM leads to a much stronger suppression in the matter power spectrum at late-times, and therefore to a more significant decrease in $\sigma_8$. The impact of the $\nu_{\rm NT}\Lambda$CDM is barely visible on the BAO scale and luminosity distance, but does affect $f\sigma_8$ predictions. The model is therefore further constrained by BOSS redshift space distortions data. Future measurements of the matter power-spectrum and $f\sigma_8$ at late times will further test this scenario
\cite{Benisty:2020kdt}.

  Although the MCMC analysis is carried out for sterile particles with the above-described momentum distributions, it has
implications for a wide class of models. As is well known (see  e.g. \cite{acero, Cuoco:2005qr}), the cosmological implications of a hot a sterile component 
is captured effectively by just two parameters: a) the contribution of the component to the present-day energy density, 
usually reported in terms of the effective mass parameter $m_{\rm eff}$; b) the contribution of the component to 
the energy density at the time of the CMB decoupling, usually reported in terms of $\Delta N_{\rm eff}$ \footnote{In the models we will discuss,
this is same as $\Delta N_{\rm eff}$ at the time of neutrino decoupling.}. These parameters are
determined by the first two moments of the momentum distribution and the mass of the sterile particle. Two models with
equal values of $m_{\rm eff}$ and $\Delta N_{\rm eff}$ will have the same phenomenological effects even if the form of the momentum distribution is different. 
We use these properties to translate the  results of the analysis for our model  parameters to results on the effective parameters. 
Our results, therefore, have direct implications for other well-motivated momentum distributions  such as  a thermal distribution with a different temperature from that of the Standard Model\cite{Feng:2008mu, Das:2010ts,Das:2017nub,Berlin:2019pbq}, the Dodelson-Widrow  distribution \cite{Dodelson:1993je}  or distributions similar to the Dodeldon-Widrow discussed in Refs~.\cite{sim1, sim2}.

Our paper is structured as follows: In Sec.~\ref{sec:HDM}, we present our model and the mapping onto generic phenomenological parameters; in Sec.~\ref{sec:mcmc}, we perform an MCMC analysis against a suite of up-to-date cosmological data and discuss the extent to which the $\nu_{\rm NT}\Lambda$CDM can resolve the $S_8$-tension; in Sec.~\ref{disc}, we draw implications of our results for other HDM models; finally, we conclude in Sec.~\ref{sec:concl}.

\section{Non-thermal hot dark matter}
\label{sec:HDM}
\subsection{The model}
\label{model}
The physics 
 of a constituent species of dark matter depends on its mass, interactions and also on its momentum distribution function. For species that thermalize, the process of thermalization brings the momentum distribution to the Fermi-Dirac or Bose-Einstein form. On the other hand, for non-thermal
constituents the momentum distribution is determined by their production   mechanism. Thus, it is important
to isolate natural production mechanisms for species that can constitute the dark matter, the associated momentum distribution and their implications for cosmology.

  In this section, we will review the basics of the production mechanism and the form of the momentum distribution that we will be considering. Our discussion
 will be brief, we refer the reader to Ref.~\cite{decayy} and the references therein for details. At early times, the 
 energy density of the universe will be taken to be dominated by cold particles of a species $\varphi$. We will
 denote the mass of the particles of $\varphi$ by $m_{\varp}$ and their decay width to be $\tau$. We will be focusing  on the
 case when the $\varp$ is the inflaton, with inflation taking place at the GUT scale and decays of the inflaton taking place due to
 a non-renormalizable interaction at the GUT scale. Thus, we take $m_{\varphi} \sim 10^{-6} M_{\rm pl}$
 and $\tau \sim 10^{8}/ m_{\varphi}$. The branching ratio of the
  $\varp$ particles to the sterile particles will be taken to be $B_{\rm sp}$, the sterile particles so produced will be
  taken not to thermalize. We will assume that the other decay products thermalize, as this sector would contain the
  Standard Model, we will refer to it as the Standard Model sector. All decay products will be taken to be relativistic
  at the time of production. As the $\varp$ particles decay, the universe goes into a matter to radiation epoch, finally 
  becoming fully radiation-dominated.
  
    During the matter to radiation-dominated epoch the evolution of the universe is governed by the equations
\be
\label{matcon}
  \dot{\rho}_{\rm mat} + 3 H \rho_{\rm mat} = - { \rho_{\rm mat} \over \tau},
\ee
\be
\label{radcon}
  \dot{\rho}_{\rm rad} + 4H \rho_{\rm rad} = + {\rho_{\rm mat} \over \tau},
\ee
and
\be
\label{evo}
    H = \left( { \dot{a} \over a } \right) = \sqrt{ { { \rho_{\rm mat} + \rho_{\rm rad} } \over 3 M^2_{\rm pl} }}.
\ee    
 In the above,  $\rho_{\rm mat}$ denotes the energy density in the matter and $\rho_{\rm rad}$ is the energy density
 in radiation. The energy density in radiation is  the sum of the energy densities in the Standard Model sector and the sterile particles (since the
 sterile particles are highly relativistic at the time of production, they contribute to the energy density as radiation when decays take place). It is useful to introduce  dimensionless variables.
 \begin{eqnarray}
 \label{sdef}
  &   \theta = { t \over \tau}, \ \ \
   \s(\theta) = a (\tau \theta), \ \ \
   \nonumber \\
 &    e_{\rm mat}(\theta) = { \tau^2 \rho_{\rm mat} (\tau \theta) \over M_{\rm pl}^2 } \ \ \ \textrm{and} \ \ \
  e_{\rm rad}(\theta) = { \tau^2 \rho_{\rm rad}  (\tau \theta) \over M_{\rm pl}^2 }. \ \ \ 
 \end{eqnarray}

 Once almost all  $\varphi$ particles have decayed, one can take  the universe to be
 composed of a thermal bath (which contains the Standard Model sector)  and the sterile particles
governed by the standard cosmological evolution equations. In practice, we will start with a matter-dominated
universe at an `initial time' $(t = \theta =0)$, evolve the universe using 
the equations \pref{matcon}, \pref{radcon} and \pref{evo} up to a fiducial dimensionless time $\theta^{*}$ which is large enough
so that almost the $\varphi$ particles have decayed by that time (we will choose $\theta^{*} = 15$ in practice). We use the results
of this procedure as initial conditions for the standard cosmological evolution. For the initial energy densities, we choose
$e_{\rm mat}(0) = {4\over3} \alpha$ and $e_{\rm rad}(0) =0 $, with $\alpha \gg 1$ (the factor of 4/3 is included as it leads to some simplifications
in the equations, for the numerical application we  take $\alpha = 10^{4}$). This implies that initially the universe is  completely matter-dominated, 
with the initial Hubble $(H_{\rm in})$ satisfying $H_{\rm in} \tau \gg 1$. This ensures that our results are independent of the choice of initial conditions.

    The momentum distribution of the sterile particles can be computed from the fact that, as a result of the decays, the co-moving number density of the  sterile particles falls off as $N(t) = N(0) e^{-t/ \tau}$
 (with the branching ratio to the sterile particles being $B_{\rm sp}$), and once produced
 the sterile particles free-stream. We will be making use of publicly available package CLASS \cite{class101, class102, class104} to incorporate the effects of the sterile particles, which takes as input the momentum distribution of the sterile particles today. 
 This was obtained in \cite{decayy} to be
 \be
 \label{fq}
   f(\vec{q}) =  {  32 \over { \pi \hat{E}^3 } } \left( { N(0) B_{\rm sp}  \over { \hat{s}^{3}(\theta^{*} ) } } \right)
     { e^{- \hat{s}^{-1}(y)} \over |\vec{q}|^{3} \hat{H}(\hat{s}^{-1}(y))},
 \ee
  where 
  \be
  \label{yyyfd}
     y = { |\vec{q}| \over 4} \hat{s}(\theta^*),
  \ee
  and the argument of the  function $\vec{q}$ is constrained so that
\be
\label{qran}
   {4 \over \hat{s}(\theta^{*}) } < |\vec{q}| < 4.
  \ee
where $\hat{E} = m_{\varphi}/2$, $N(0)$ is the initial number density  of the $\varphi$ particles, $\hat{s}(\theta^*)$ the 
value scale factor\footnote{In \pref{fq} the convention is that the scale factor is 1 at $\theta = 0$} at the fiducial dimensionless time $\theta^{*}$,  $\hat{s}^{-1}$ the functional inverse of the 
scale factor function as a function of the dimensionless time  and $\hat{H} = \hat{s}'(\theta) /  \hat{s}(\theta)$ the dimensionless Hubble constant. The momentum $\vec{q}$ in \pref{fq} is the momentum in units of the typical momentum magnitude of the sterile particles today $(T_{\rm ncdm,0})$. The typical momentum magnitude was found to be 
\bel{tmomf}
  T_{\rm ncdm,0}  =  0.418 \left( \mv^2 \tau \over M_{\rm pl} \right)^{1/2} { T_{\rm cmb}  \over {(1 - B_{\rm sp})^{1/4} }} \equiv \zeta T_{\rm cmb}
\ee
 in \cite{decayy}. The distribution function in \pref{fq} is in units of $T_{\rm ncdm,0}^{3}$. Thus $f(\vec{q}) d^3{q}$ gives the number density of particles with  their dimensionless momentum in the interval $(q_i, q_i + dq_i)$ with the number density measured in units of $T^{3}_{\rm ncdm,0}$.
 
\begin{figure}[h]
\centering
\includegraphics[width=0.4 \textwidth]{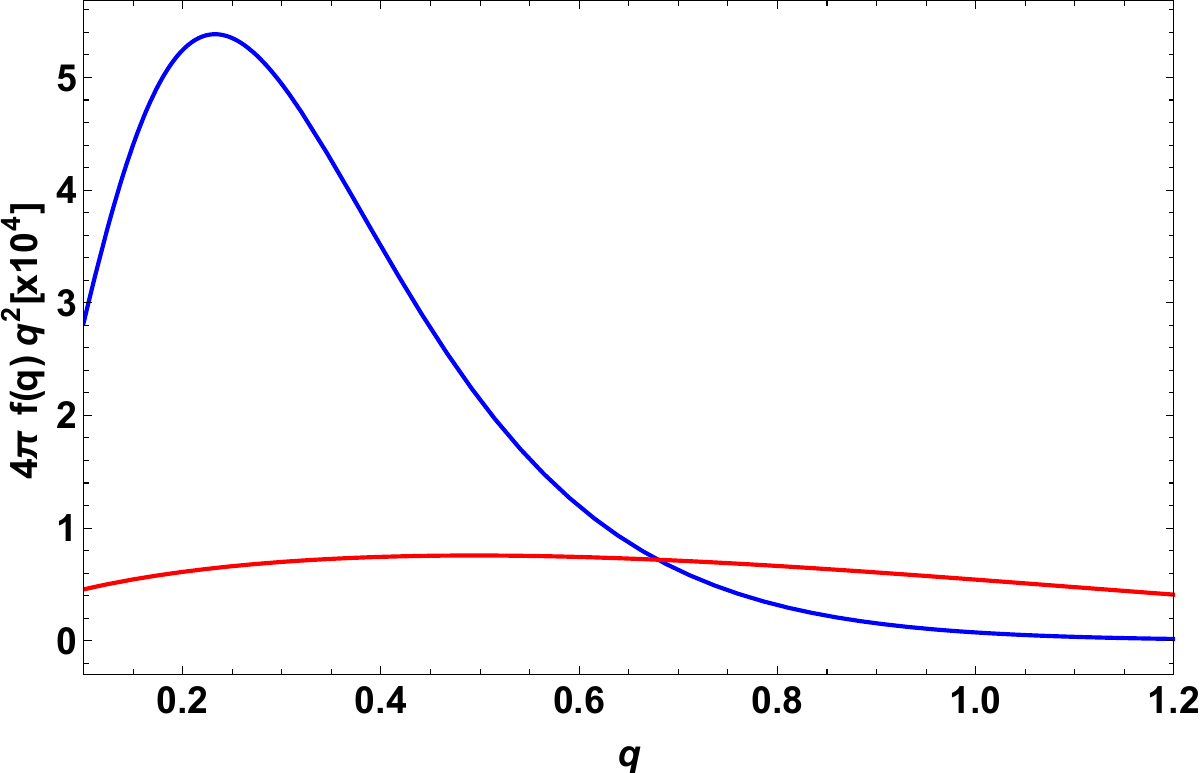}
\caption{Comparative plot  with a thermal distribution. The nonthermal distribution is plotted
in orange and is for value of the parameter $ m_{\varphi}= 10^{-6} M_{\rm pl} $, $\tau = {10^{8} / m_{\varphi}}$ . The thermal 
distribution is in blue. The momenta  and the distribution functions for
both plots are in units of $T_{\rm ncdm,0}$ as associated with the above values of $m_{\varphi}$ and $\tau$. $\Delta N_{\rm eff}$ is taken to be 0.15 for both distributions. The x axis label ${q} \equiv |\vec{q}|$
as defined in the main text.}
\label{compareTh}
\end{figure}

 Note that although naively $f(\vec{q})$ seems to depend
 on $N(0)$, the full expression is independent of $N(0)$ as long as we take the universe to be completely matter-dominated
 at the initial time. It is interesting to compare the distribution to a thermal one, as shown in figure \ref{compareTh}. We focus on the range $q \equiv |\vec{q}| \in[0.1,1.2]$ because the distribution falls off beyond that range \cite{decayy}.
 For the same value of $\Delta N_{\rm eff}$, the non-thermal distribution is 
peaked at higher values of the momentum, but is much broader. 
The mean momentum of sterile particles is greater than that of the CMB by the factor $\zeta$ defined in \pref{tmomf}. For our choice
 of parameters $\zeta \sim 5$. The sterile particles become non-relativistic when their typical momentum becomes of the order of their mass i.e
 the temperature of the Standard Model plasma becomes of the order {\color{red} ${m_{\rm sp} / 5}$}.
 
\subsection{Mapping onto generic parameters}
\label{macpi}

   Our model has four microscopic  parameters:  $m_{\varphi}, \tau$ (the mass and lifetime of the decaying particle), $B_{\rm sp}$
 (the branching ratio for decay to the sterile particle) and $m_{\rm sp}$ (the mass of the sterile particle) in addition to
 those of $\Lambda$CDM.  Our choice of the 
 first two parameters ($m_{\varphi} \sim 10^{-6} M_{\rm pl}$ and $\tau \sim 10^{8}/ m_{\varphi}$) is motivated by taking $\varphi$ to be driving inflation at the GUT scale and decaying by GUT scale interactions. 
On the other hand, the other parameters $B_{\rm sp}$ and $m_{\rm sp}$ will be traded for effective parameters more directly connected to observables.
 Indeed, the physical effects of new sterile particle/species on the cosmological background and perturbation evolution can be  completely described by three parameters:  $\Delta N_{\rm eff}$ (the effective number
  of relativistic neutrinos at the time  of neutrino decoupling), $w_{\rm sp} \equiv \Omega_{\rm sp} h^2$ (where $\Omega_{\rm sp}$ is the fractional contribution
  of the particle to today's energy density and h the reduced Hubble parameter, this is often characterised by the effective mass of the particle $\mef. = w_{\rm sp} 94.05 {\rm eV}$) and $\lambda_{\rm FS}$ (the free-streaming length associated with the species). The free-streaming length is determined once the first two quantities are known, hence effectively there are two parameters \cite{acero}. Physically, the two parameters of interest for reducing $\sigma_8$ are $w_{\rm sp}$, which fixes the depth of the power-suppression, and $\lambda_{\rm FS}$, which fixes the scale above which modes are suppressed. Still, for their simpler connection with micro-physics, here we take  $\Delta N_{\rm eff}$ and $\mef$ as two independent quantities, defined as
  
\begin{equation}
\Delta N_{\rm eff} \equiv  \frac{\rho_s^{\rm rel}}{\rho_\nu}
= {1 \over \pi^{2} } \left[
{ \int \!\! d p \,\, {p}^{3} \hat{f}({p})}
\right] /
\left[{\frac{7}{8} \frac{\pi^2}{15}
{T_\nu^{\rm id}}^4} \right]  
\end{equation}
with $T_\nu^{\rm id} \equiv (4/11)^{1/3} T_{\gamma}$ and
\begin{equation}
\frac{\mef}{94.05 {\rm eV}} \equiv \omega_s \equiv \Omega_s h^2 = {1 \over \pi^{2}}\left[
{{m_{\rm sp}} \int \!\! d p \,\, p^{2} \hat{f}({p})} 
\right] \times \left[\frac{h^2}{\rho_c^0}\right]
\end{equation}

 where $\hat{f}(p)$  is  the distribution function as a function of the magnitude of the 
physical momentum in the conventions of \cite{acero} .\footnote{In these conventions, an
additional species of neutrinos at temperature $T_s$ has $\hat{f}(p) = {1 \over {e^{p/T_s} +1}}$. For our non-thermal
distribution  $\hat{f}(p) = 4 \pi^3 f({p \over T_{\rm ncdm,0}}  \hat{\bf{e}})$, where the function $f$ is as defined in \pref{fq} and $\hat{\bf{e}}$ is an arbitrary unit vector.}
$\rho_c^0$ is the critical density today and $h$ the reduced Hubble parameter. The
 In our model, the effective parameters $m_{\rm eff}$ and $\Delta N_{\rm eff}$ in terms of the microscopic parameters are given by \cite{decayy}:
 \begin{equation}
 \label{delN}
 \Delta N_{\rm eff} =  {43 \over 7} { B_{\rm sp} \over {1 -  B_{\rm sp}} } 
 \left( { g_{*} (T(t_{\nu}) ) \over g_{*} (T(t^{*})) } \right)^{1/3}
 \end{equation}
and
 \begin{equation}
 \label{meff}
   m^{\rm eff}_{\rm sp}     =    { 62.1 m_{\rm sp} \over {  g_{*}^{1/4} (T(t^{*})) } }   { B_{\rm sp} \over {(1 - B_{\rm sp})^{3/4}}}  \left( {\mpl \over \tau \mv^2} \right)^{1/2},
 \end{equation}
 where $g_{*} (T(t_{\nu}))$ and $g_{*} (T(t^{*}))$ are the effective number of degrees of freedom at the time of neutrino
 decoupling and the end of the reheating epoch (we will take the latter to be equal to $100$). 
 We will thus scan over $m_{\rm sp}$ and $B_{\rm sp}$ (keeping  $m_{\varphi} = 10^{-6} M_{\rm pl}$
 and $\tau \sim 10^{8}/ m_{\varphi}$ fixed), and use Eqs.~\ref{meff} and \ref{delN} to relate to phenomenological parameters. 
 See Appendix B for a discussion of a model with a different values of $m_{\varphi}$ and $\tau$ and it matches with the above scan
 when the effective parameters match.
 In sec.~\ref{disc}, we will then translate our results for two other models of interest. 
  Note that the free-streaming length can be simply extracted from our analysis using the relation \cite{Lesgourgues:2006nd}:
 \begin{equation}
     \lambda_{\rm fs}(t) = 2\pi\sqrt{\frac{2}{3}}\frac{v_{\rm s}(t)}{H(t)}\,,
 \end{equation}
 where, when evaluated today at $t=t_0$, one has  \cite{acero}
  \begin{equation}
v_{\rm s}(t_0)\simeq 5.618\times10^{-6}\frac{\Delta N_{\rm eff}}{\omega_s}\,.
  \end{equation}

%

\begin{table*}[t]
    \centering
    \scalebox{0.75}{
    \begin{tabular}{|l|c|c|c|c|c|c|} 
    \hline
    Model & \multicolumn{2}{|c|}{$\Lambda$CDM}& \multicolumn{2}{|c|}{$\nu\Lambda$CDM} & 
\multicolumn{2}{|c|}{$\nu_{\rm NT}\Lambda$CDM}  \\
\hline 
     \hline Parameter&Planck & Planck + $S_8$ &Planck & Planck + $S_8$ & Planck & Planck + $S_8$\\ \hline         
      $100~\omega_{b}$ & $2.24(2.24)_{-0.015}^{+0.014}$& $2.252( 2.256)_{-0.015}^{+0.014}$ & $2.244_{-0.018}^{+0.016}$ & $2.257(2.259)\pm0.017$& $2.241(2.247)_{-0.016}^{+0.015}$ &$2.247(2.247)_{-0.015}^{+0.014}$ \\
     $\omega_{\rm cdm}$ & $0.1198(0.1195)_{-0.0012}^{+0.0013}$& $0.1182(0 .1177)\pm0.0011$ & $0.1217_{-0.002}^{+0.0015}$& $0.1198(0.1182)_{-0.0018}^{+0.0013}$& $0.118(0.1198)_{-0.0022}^{+0.0041}$& $0.1142(0.1110)_{-0.003}^{+0.0049}$\\
     $100*\theta_{s }$ &$1.04190(1.04178)_{-0.0003}^{+0.00029}$ &  $ 1.04202(1.04217)_{-0.0003}^{+0.00029}$& $1.04166_{-0.00033}^{+0.00037}$&  $1.04179(1.04191)_{-0.00032}^{+0.00035}$ & $1.04180(1.04187)\pm0.00032$  &  $1.04186( 1.04190)_{-0.00029}^{+0.00031}$\\
     $n_{s }$ & $0.9661(0.9663)_{-0.0043}^{+0.0041}$&$0.9695(0.971)_{-0.0041}^{+0.0039}$ &$0.9685_{-0.006}^{+0.0049}$ &$0.9717(0.9732)_{-0.0056}^{+0.0048}$ & $0.9652( 0.9677)_{-0.0051}^{+0.0044}$  &$0.9652(0.9661)_{-0.0045}^{+0.0047}$ \\
     ${\rm ln}(10^{10}A_{s })$ & $3.044(3.044)\pm0.014$  & $3.041(3.042)_{-0.015}^{+0.014}$& $3.052_{-0.016}^{+0.015}$& $3.048(3.050)_{-0.017}^{+0.016}$ & $3.047
     (3.0480)\pm0.015$& $3.046(3.044)_{-0.016}^{+0.014}$\\
     $\tau_{\rm reio }$ & $0.0541(0.0541)_{-0.0071}^{+0.0075}$&  $0.0542(0.0556)_{-0.0078}^{+0.0074}$& $0.0558_{-0.0081}^{+0.0073}$& $0.0555( 0.0590)_{-0.0082}^{+0.0077}$ &  $0.0545(0.0559)_{-0.0081}^{+0.0073}$ &$0.0548(0.0536)_{-0.0079}^{+0.0069}$  \\
     $m_\nu$ [eV] & $-$ &$-$ & $ <0.073$& $<0.1(0)$  & $-$  & $-$ \\
     $\mef$ [eV]  & $-$  & $-$&$-$ & $-$ &  $<1.02 (0)$ &$0.67(0.90)_{-0.48}^{+0.26}$  \\
    $\nef$ &  $-$  &$-$  & $<0.28$ &  $<0.24(0.03)$& $< 0.15(0.03)$& $0.0614(0.034)_{-0.047}^{+0.0052}$  \\

    \hline 
    $S_{8 }$ & $0.834(0.832)\pm0.013$&  $0.814(0.809)_{-0.011}^{+0.01}$& $0.834(0.838)_{-0.013}^{+0.013}$  & $0.812(0.814)\pm0.011$ & $0.815(0.831)_{-0.018}^{+0.022}$  &$0.789(0.791)\pm0.016$ \\
    $\Omega_{m }$ & $0.3078(0.3068)_{-0.0076}^{+0.0074}$&$0.2981(0.2948)_{-0.0066}^{+0.0061}$ & $0.3154(0.3084)_{-0.015}^{+0.0094}$&$0.3084(0.295)_{-0.018}^{+0.0081}$ & $0.3138(0.305)_{-0.0097}^{+0.0084}$  & $0.311(0.308)_{-0.01}^{+0.008}$\\ 
    $H_0$ [km/s/Mpc] &$68(68.04)\pm0.56$ & $68.73(68.99)_{-0.51}^{+0.49}$& $67.83( 67.95)_{-1}^{+1.2}$  &$68.26(69.11)_{-0.93}^{+1.5}$ & $67.72(68.34)_{-0.65}^{+0.62}$ &$67.91(68.04)_{-0.61}^{+0.67}$ \\
        \hline 
$\chi^2_{\rm min}$ & 2774.8 & 2783.4& 2774.9 & 2782.0 & 2775.0&  2778.60 \\ 
\hline
    \end{tabular}
    }
    \caption{ The mean (best-fit) $\pm1\sigma$ error of the cosmological parameters in the $\Lambda$CDM and $\nu_{\rm NT}\Lambda$CDM model obtained from the analysis of Planck \cite{Aghanim:2018eyx} and Planck+$S_8$ \cite{Heymans:2020gsg} data. 
    The definition of $\mef$ is given in Eq.~\ref{meff}.
    Upper limits are given at the 95\% C.L. }
    \label{tab:MCMC1}
\end{table*}

\begin{table*}[t]
    \centering
    \scalebox{0.75}{
    \begin{tabular}{|l|c|c|c|c|c|c|} 
    \hline
    Model & \multicolumn{2}{|c|}{$\Lambda$CDM}& \multicolumn{2}{|c|}{$\nu\Lambda$CDM} & 
\multicolumn{2}{|c|}{$\nu_{\rm NT}\Lambda$CDM}  \\
\hline 
     \hline Parameter & Planck+Ext & Planck+Ext+$S_8$ & Planck+Ext & Planck+Ext+$S_8$ & Planck+Ext & Planck+Ext+$S_8$\\ \hline         
     $100~\omega_{b}$ &$2.241(2.238)_{-0.014}^{+0.013}$  & $2.248(2.258)\pm{0.013}$&   $2.249(2.248)\pm0.015$  &
     $2.257(2.250)\pm0.015$&$2.245(2.245)\pm0.014$ & $2.250(2.253)_{-0.014}^{+0.013}$\\
     $\omega_{\rm cdm}$ & $0.1197(0.1204)\pm0.0009$ &  $0.1187(0.1182)_{-0.0008}^{+0.0009}$ & $0.121(0.1194)_{-0.0019}^{+0.0012}$ &
     $0.1198(0.1186)_{-0.0017}^{+0.0011}$ &$0.1181(0.1179)_{-0.0018}^{+0.0030}$ & $0.1152(0.1101)_{-0.0023}^{+0.0036}$  \\
     $100*\theta_{s }$ &  $1.04192(1.04204)_{-0.00029}^{+0.00028}$& $ 1.04197(1.04186)_{-0.00029}^{+0.0003}$ & $1.04172(1.04194)_{-0.00031}^{+0.00034}$ &  $1.04179( 1.04194)_{-0.00031}^{+0.00036}$ & $1.04187(1.04193)_{-0.00029}^{+0.0003}$&$1.04193(1.04194)_{-0.00028}^{+0.00029}$ \\
     $n_{s }$ &  $0.9664(0.9660)_{-0.0037}^{+0.0038}$&  $0.9683(0.9705)_{-0.0038}^{+0.0036}$&$0.9699(0.9693)_{-0.0049}^{+0.0044}$  &$0.9721(0.9706)_{-0.0048}^{+0.0043}$  &  $0.9667(0.9664)_{-0.0041}^{+0.0039}$ &$0.9669(0.9678)_{-0.004}^{+0.0039}$ \\
     ${\rm ln}(10^{10}A_{s })$ & $3.044(3.05)_{-0.015}^{+0.014}$ & $3.038(3.045)_{-0.015}^{+0.013}$&  $3.052(3.049)_{-0.016}^{+0.014}$ & 
     $3.046(3.035)_{-0.016}^{+0.015}$&$3.049(3.052)_{-0.015}^{+0.014}$ &$3.046(3.054)_{-0.015}^{+0.014}$ \\
     $\tau_{\rm reio }$ &  $0.0542(0.0574)_{-0.0073}^{+0.0069}$&$0.0526(0.056)_{-0.0076}^{+0.0069}$ & $0.0561(0.0569)_{-0.0081}^{+0.0066}$  & 
     $0.0548(0.0515)_{-0.0081}^{+0.0073}$ &$0.0559(0.0576)_{-0.0076}^{+0.007}$ & $0.0556(0.0586)_{-0.0076}^{+0.0068}$\\
     $m_\nu$ [eV] & $-$& $-$ &  $<0.040(0.005)$ &  $<0.057(0.01)$ & $-$  & $-$ \\
     $\mef$ [eV]  &$-$ & $-$ &  $-$ & $-$ & $<0.67(0.21)$ & $0.48(0.92)_{-0.36}^{+0.17}$\\
     $\nef$ &  $-$  &$-$  & $<0.27(0.02)$& $<0.26(0.006)$ & $< 0.12(0.02)$&   $0.0457(0.0336)_{-0.031}^{+0.0038}$ \\
    \hline 
    $S_{8 }$ &  $0.832(0.842)\pm0.011$&  $0.818(0.815)_{-0.0094}^{+0.0091}$& $0.830(0.827)\pm0.011$ & $0.814( 0.815)_{-0.0097}^{+0.01}$ & $0.815(0.820)_{-0.015}^{+0.017}$&$0.795(0.787)_{-0.013}^{+0.015}$  \\
    $\Omega_{m }$ & $0.3067(0.31)\pm0.0055$ & $0.3007(0.2974)_{-0.0049}^{+0.0051}$& $0.3084(0.3042)_{-0.006}^{+0.0059}$ & $0.3045(0.3037)_{-0.0072}^{+0.0061}$ & $0.309(0.308)_{-0.0061}^{+0.0057}$& $0.306(0.304)\pm0.006$\\ 
    $H_0$ [km/s/Mpc] & $68.07(67.82)_{-0.43}^{+0.41}$ &  $68.52(68.78)_{-0.4}^{+0.38}$& $68.35(68.33)_{-0.7}^{+0.56}$ & $68.58(68.28)_{-0.73}^{+0.64}$& $68.06(67.97)_{-0.47}^{+0.44}$  &  $68.22(68.37)_{-0.43}^{+0.41}$ \\
    \hline 
    $\chi^2_{\rm min}$ & 3810.4 & 3818.2&   3809.5 &3816.4 & 3809.7 & 3814.5  \\
    \hline 
    \end{tabular}
    }
    \caption{Same as Tab.\ref{tab:MCMC1} with the addition of 'Ext' data, which refers to the combination BAO/FS+Pantheon.}
    \label{tab:MCMC2}
\end{table*}

\section{Resolving the $S_8$ tension with a non-thermal sterile neutrino}
\label{sec:mcmc}
\subsection{Details of the analysis}
We perform a comprehensive MonteCarlo Markov Chain (MCMC) analysis and  confront the non-thermal hot dark matter model to various combination of the following data sets:
\begin{itemize}
    
    \item Planck 2018 measurements of the low-$\ell$ CMB TT, EE, and  high-$\ell$ TT, TE, EE power spectra, together with the gravitational lensing potential reconstruction \cite{Aghanim:2018eyx}. 
        
    \item The BAO measurements from 6dFGS at $z=0.106$~\cite{Beutler:2011hx}, SDSS DR7 at $z=0.15$~\cite{Ross:2014qpa}, BOSS DR12 at $z=0.38, 0.51$ and $0.61$~\cite{Alam:2016hwk}, and the joint constraints from eBOSS DR14 Ly-$\alpha$ auto-correlation at $z=2.34$~\cite{Agathe:2019vsu} and cross-correlation at $z=2.35$~\cite{Blomqvist:2019rah}.
    
    \item The measurements of the growth function $f\sigma_8(z)$ (FS) from the CMASS and LOWZ galaxy samples of BOSS DR12 at $z = 0.38$, $0.51$, and $0.61$~\cite{Alam:2016hwk}.
    
    \item The Pantheon SNIa catalogue, spanning redshifts $0.01 < z < 2.3$~\cite{Scolnic:2017caz}.
    
    \item The KIDS1000+BOSS+2dfLenS weak lensing data, compressed as a a split-normal likelihood on the parameter $S_8=0.766^{+0.02}_{-0.014}$~\cite{Heymans:2020gsg}.

\end{itemize}

Our baseline cosmology consists in the following combination of the six $\Lambda$CDM parameters $\{\omega_b,\omega_{\rm cdm},100\times\theta_s,n_s,{\rm ln}(10^{10}A_s),\tau_{\rm reio}\}$, plus two parameters describing the non-thermal hot dark matter, namely $\{m_{\rm sp}, B_{\rm sp}\}$. We dub this model $\nu_{\rm NT}\Lambda$CDM. Standard model neutrinos are assumed to be massless.

To better understand how the $\nu_{\rm NT}\Lambda$CDM model can resolve the $S_8$ tension, we will compare it to the standard $\Lambda$CDM model with massless neutrinos, as well as to the $\Lambda$CDM model with free neutrino masses $m_\nu$ and additional relativistic degrees of freedom $\nef$.  In that latter case, we assume degenerate neutrino masses and a free-streaming $\nef$. Note that in this model, the $\nef$ component does not become massive at late times, contrary to what happens in the non-thermal neutrino model. This will play a key role in the difference between the two models. We dub this model $\nu\Lambda$CDM.  We run our MCMCs with the Metropolis-Hasting algorithm as implemented in the MontePython-v3 \cite{Brinckmann:2018cvx} code interfaced with our modified version of CLASS. All reported $\chi^2_{\rm min}$ are obtained with the python package {\sc iMinuit \footnote{\url{https://iminuit.readthedocs.io/}}} \cite{James:1975dr}. We make use of a Choleski decomposition to better handle a large number of nuisance parameters \cite{Lewis:1999bs} and consider chains to be converged with the Gelman-Rubin convergence criterium $R-1\lesssim0.05$ \cite{Gelman:1992zz}.

\begin{figure*}[t]
\centering
\includegraphics[scale=0.3]{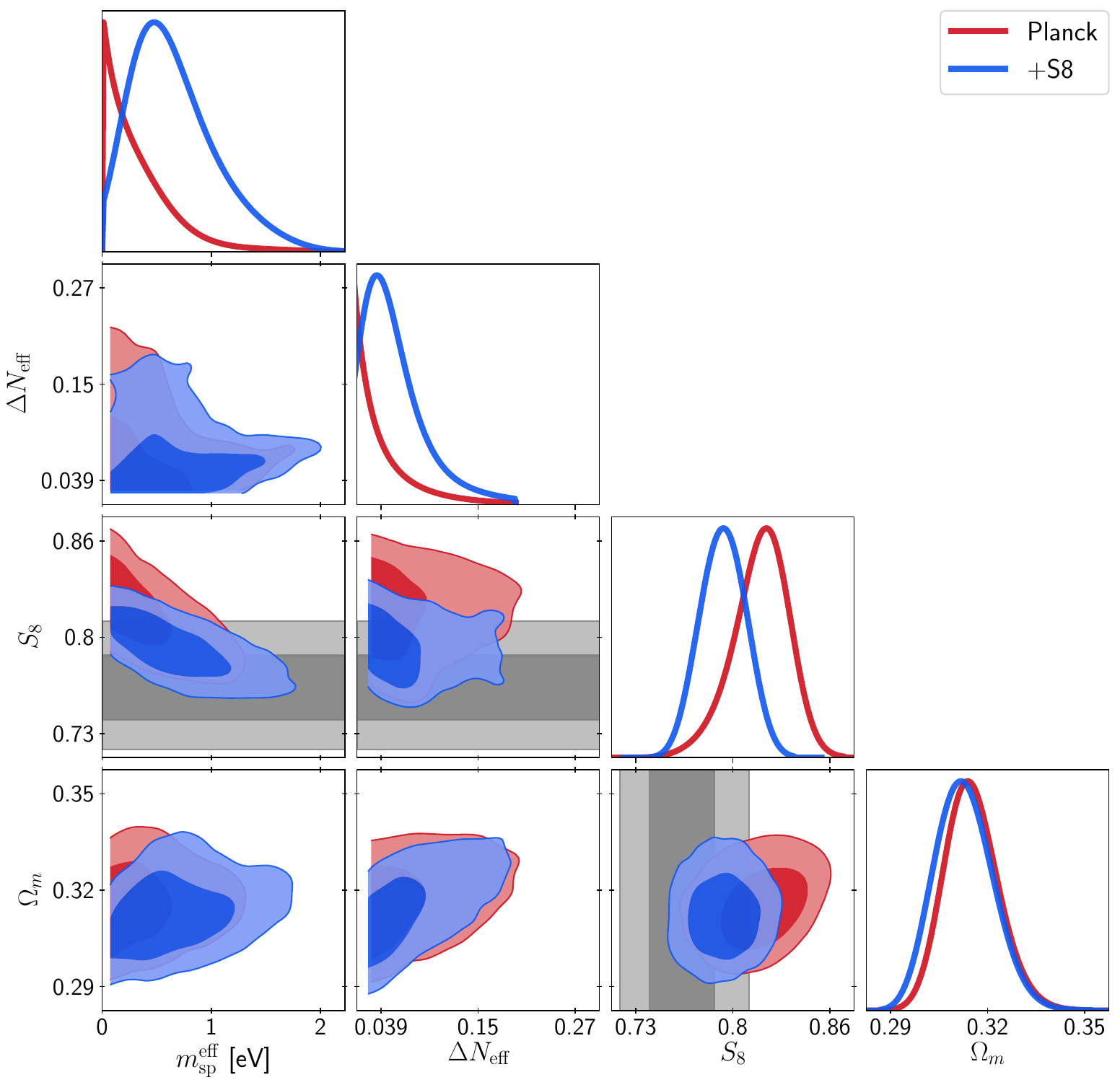}
\includegraphics[scale=0.3]{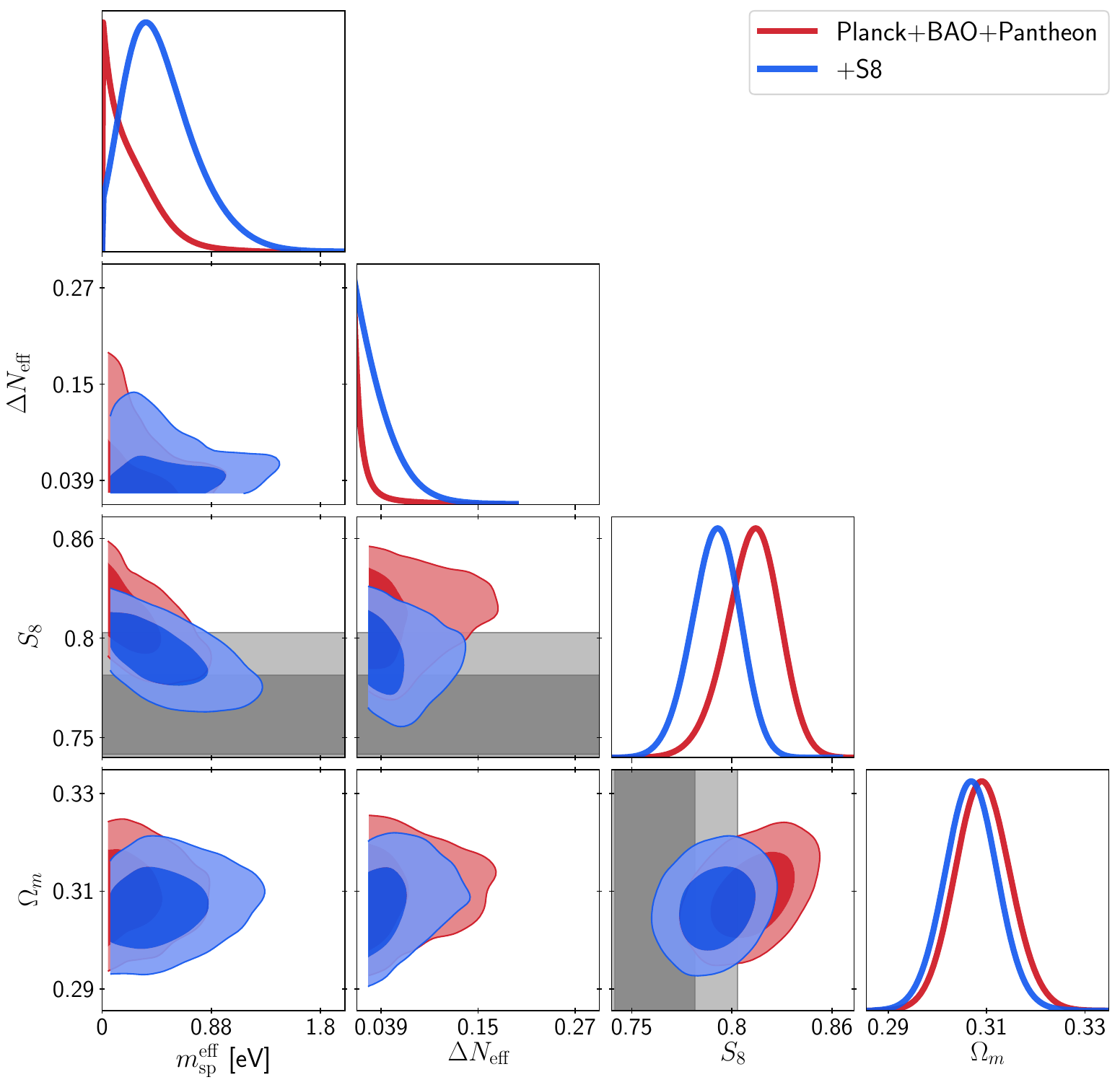}
\caption{Reconstructed 2D posterior distributions of $\{\mef,\nef, S_8, \Omega_m\}$ with Planck and Planck+$S_8$ data (left panel) or Planck+BAO+SN1a and  Planck+BAO+SN1a+$S_8$ data (right panel). \label{fig:MCMC_NT}}
\end{figure*}

\begin{figure*}[t]
\centering

\includegraphics[scale=0.3]{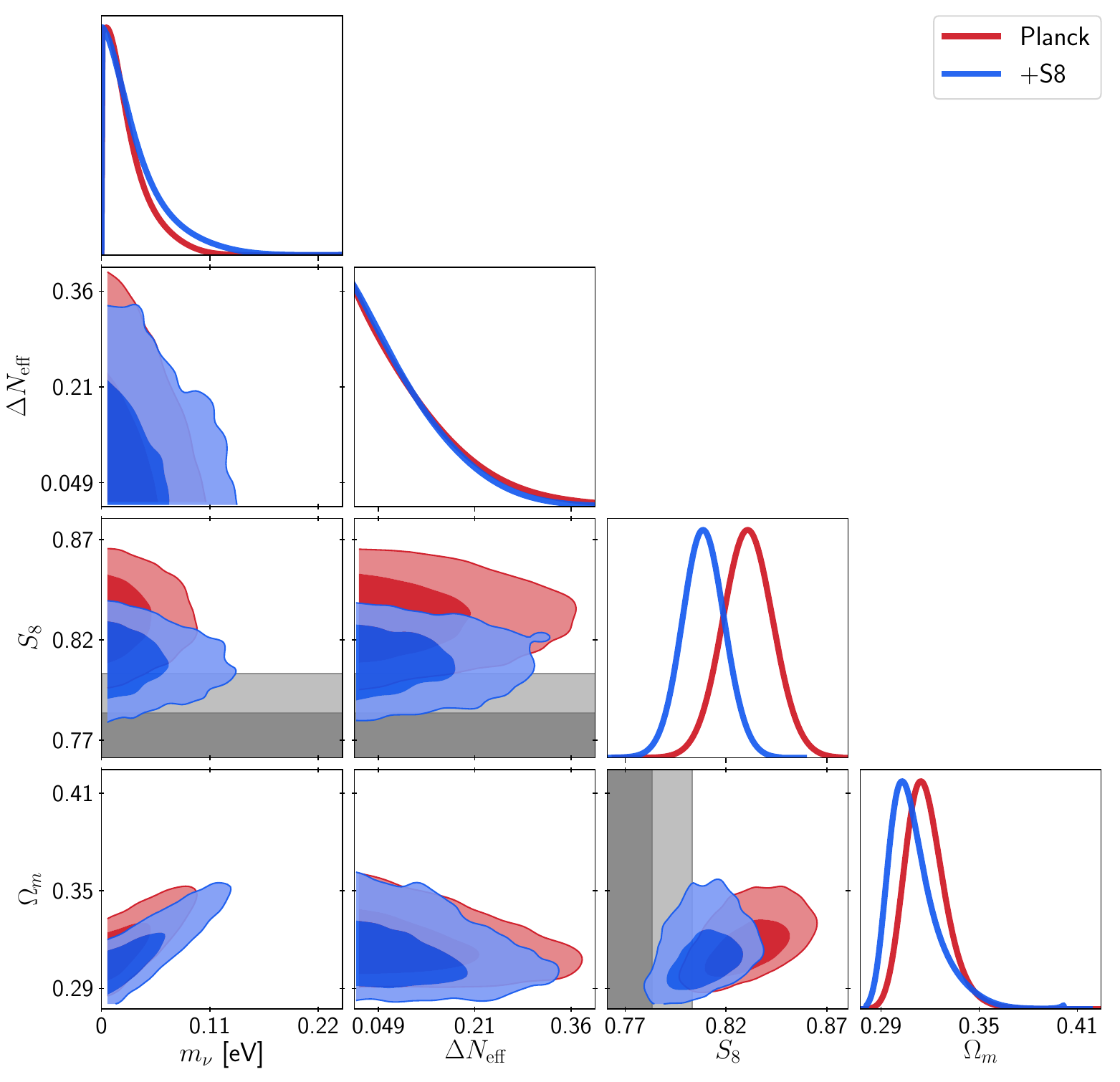}
\includegraphics[scale=0.3]{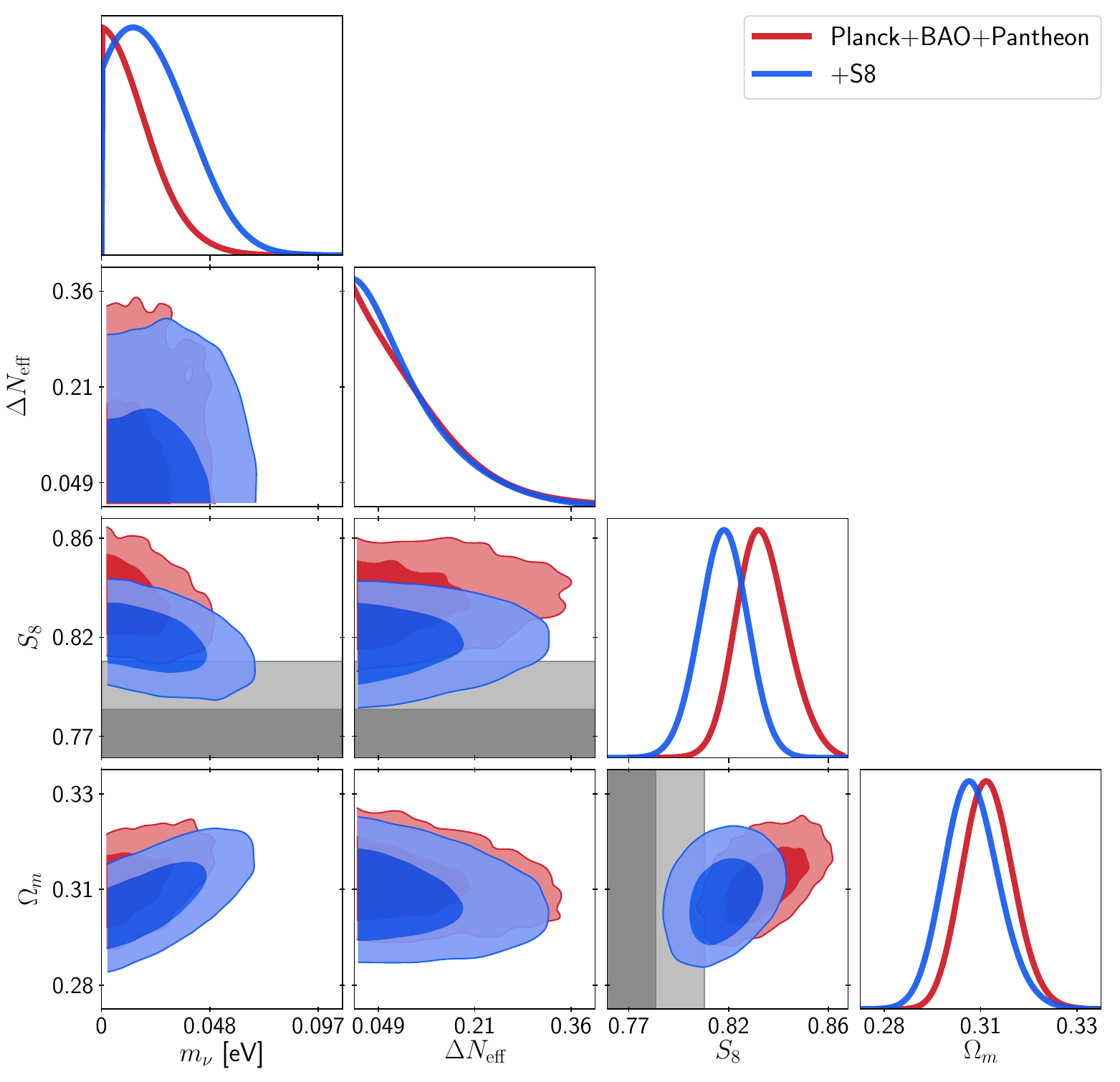}
\caption{Same as fig.~\ref{fig:MCMC_NT} in the thermal neutrino case. \label{fig:MCMC_T}}
\end{figure*}

\subsection{Results }

We run two sets of analysis; in the first one, we confront the $\Lambda$CDM, $\nu\Lambda$CDM and $\nu_{\rm NT}\Lambda$CDM models to Planck only and Planck+$S_8$. In the second one, we add the BAO and Pantheon data to our analysis.
Our main results are reported in Tabs.~\ref{tab:MCMC1} and \ref{tab:MCMC2} and displayed on Figs.~\ref{fig:MCMC_NT} and \ref{fig:MCMC_T}. We report results in the $\nu_{\rm NT}\Lambda$CDM in terms of $\Delta N_{\rm eff}$ and $m_{\rm sp}^{\rm eff}$ defined in Eqs.~\ref{delN} and \ref{meff}. We give the $\chi^2_{\rm min}$ per experiment\footnote{ Note that we model neutrinos as degenerate in the thermal and non-thermal case, while the $\Lambda$CDM model has 2 massless, 1 massive neutrino with $m = 0.06$ eV (following Planck convention). This leads to very small differences in practice, and explain why we cannot recover exactly the $\Lambda$CDM model $\chi^2$ in the massive neutrino cases. Similarly, the non-thermal model does not 'exactly' reduce to the thermal model in some part of the parameter space. Small $\chi^2$ differences are therefore expected, and safe given their statistical insignificance.} in App.~\ref{app:chi2}.

\subsubsection{Planck only}

When the $\nu_{\rm NT}\Lambda$CDM model is confronted to Planck only, we obtain a bound\footnote{Hereinafter, we quote 1-sided constraints at 95\%C.L., and two-sided ones at 68\%C.L.}  on the mass $ \mef < 1.02$ eV and $\nef < 0.15$. Similarly, in the $\nu\Lambda$CDM case we obtain $m_\nu< 0.073 $ eV and $\nef < 0.28 $ (recall that this limit applies to individual neutrino masses in the degenerate case). The $\chi^2_{\rm min}$ of Planck in the $\nu\Lambda$CDM  and $\nu_{\rm NT}\Lambda$CDM scenario is not improved over that of $\Lambda$CDM. We note that the $\nu_{\rm NT}\Lambda$CDM model predicts a lower $S_8$ value than other models. Indeed, we find $S_8(\nu\Lambda{\rm CDM})=0.831_{-0.013}^{+0.012}$ and $S_8(\Lambda{\rm CDM})= 0.832\pm0.011$, to be compared to $S_8(\nu_{\rm NT}\Lambda{\rm CDM})=0.816_{-0.016}^{+0.022}$, i.e., a $\gtrsim1\sigma$ downward shift. As a result, the $S_8$ tension is alleviated from the $\sim2.7\sigma$ level to the $\sim1.9\sigma$ level in the non-thermal HDM model.  We note that our constraints on $\nef$ in the non-thermal case is stronger than that reported in Ref.~\cite{Aghanim:2018eyx} (constraints are identical in the thermal case). This likely comes from the impact of running on physical parameters as opposed to phenomenological parameters when exploring the parameter space.

Including the prior on $S_8$, we notice a mild detection of non-zero $\mef=0.67_{-0.48}^{+0.26}$ eV and $\nef = 0.0614_{-0.047}^{+0.0052}$ in the $\nu_{\rm NT}\Lambda{\rm CDM}$ model, while the constraints on the thermal neutrino mass simply relaxes to $m_{\nu} < 0.1$ eV. 
This translates into a reconstructed $S_8(\nu_{\rm NT}\Lambda{\rm CDM}) = 0.789\pm0.016$  and $S_8(\nu\Lambda{\rm CDM}) = 0.812\pm0.011$,  to be compared with the baseline $S_8(\Lambda{\rm CDM}) = 0.814_{-0.011}^{+0.01}$. As a consequence, the $\chi^2_{\rm min}$ in the combined analysis is lower in the non-thermal HDM case $\Delta\chi^2_{\rm min}(\nu_{\rm NT}\Lambda{\rm CDM})=\chi^2_{\rm min}(\Lambda{\rm CDM}) - \chi^2_{\rm min}(\nu_{\rm NT}\Lambda{\rm CDM})= -4.8$ than in the thermal neutrino case $\Delta\chi^2_{\rm min}(\nu\Lambda{\rm CDM})=\chi^2_{\rm min}(\Lambda{\rm CDM}) - \chi^2_{\rm min}(\nu\Lambda{\rm CDM})= -1.4$. 
If the $S_8$ tension worsens in the future, it would be interesting to perform a more complete Bayesian analysis comparing these models. 
We notice, however, that the total $\chi^2_{\rm min}$ is much less significantly affected by the inclusion of the $S_8$ prior in the non-thermal case (+3.6) than in the thermal case (+6.9), which is encouraging and indicates that the $\nu_{\rm NT}\Lambda{\rm CDM}$ model can potentially alleviate the tension between Planck and KIDS+BOSS. It remains to be seen whether this is robust to additional data sets (and in the future it should be tested against the full KiDS and BOSS likelihoods).

 Before including external data, we comment on the possibility for non-thermal hot dark matter to resolve the Hubble tension (see e.g. \cite{Riess:2020fzl,Freedman:2020dne,DiValentino:2021izs} for a review). We find that, whether we include the $S_8$ prior or not, the value of $H_0$ is barely affected by the extra $\nef$ (in fact, even shifted slightly towards lower $H_0$ due to the well-known anti-correlation with $\mef$ \cite{Lesgourgues:2011re}). We, therefore, confirm that these models cannot be responsible for the high-$H_0$ measured with some of the local probes.
\subsubsection{Planck+BOSS+SN1a}
When the BAO/FS and SN1a data are added to the analysis, the constraints on the thermal neutrino mass  and non-thermal hot dark matter mass strengthen. 
We find $\mef<0.67$ eV and $\nef < 0.12$ in the $\nu_{\rm NT}\Lambda{\rm CDM}$ model, while we get $m_{\nu}<0.04$ eV and $\nef < 0.27$ in the thermal case. 
Still, the reconstructed $S_8$ value $S_8(\nu_{\rm NT}\Lambda{\rm CDM}) = 0.814_{-0.014}^{+0.017}$ and  $S_{8 }(\nu\Lambda{\rm CDM})=0.83\pm0.011$ are slightly smaller than in the Planck-only analysis. This is because the reconstructed value of $\omega_{\rm cdm}$ is slightly smaller in the combined analysis with BAO/FS and SN1a data, regardless of the model. 

Once the prior on $S_8$ is added to the analysis, we again find a mild detection of  $m_{\rm eff}=0.48_{-0.36}^{+0.17}$
 eV and $\nef=0.0457_{-0.031}^{+0.0038}$ . 
 However, the mean value has decreased by $0.5\sigma$ due to the inclusion of BAO/FS and SN1a data. 
 This reflects in a slightly larger reconstructed $S_8$ value, $S_8(\nu_{\rm NT}\Lambda{\rm CDM}) =0.795_{-0.013}^{+0.015}$. 
 A similar pattern is observed in the thermal case, for which the relaxation of the constraint to $m_{\nu} < 0.057$ eV is much milder than without BAO/FS and SN1a data, while the reconstructed $S_8(\nu\Lambda{\rm CDM})=0.814\pm0.01$ is stable.  Looking at $\chi^2_{\rm min}$, one can see that the non-thermal case still provides a better fit $\Delta\chi^2_{\rm min}(\nu_{\rm NT}\Lambda{\rm CDM})=-3.7$ than the thermal case $\Delta\chi^2_{\rm min}(\nu\Lambda{\rm CDM})=-1.8$. However, the inclusion of the $S_8$ prior as increased the total $\chi^2_{\rm min}$ by $+4.8$ in the non-thermal case and $+6.9$ in the thermal case. 
  It is interesting to note that  the tension level between Planck and KiDS evolves from $1.9\sigma$ to $2.2\sigma$ once BAO data are included, i.e., these data worsen the tension. This is in contrast with the $\Lambda$CDM case, for which the tension goes from $2.9\sigma$ (without BAO) to $2.8\sigma$ (with BAO). More accurate BAO/FS and SN1a data could therefore pose a serious challenge to this model.

\begin{figure*}[t]
\centering

\includegraphics[width=1.5\columnwidth]{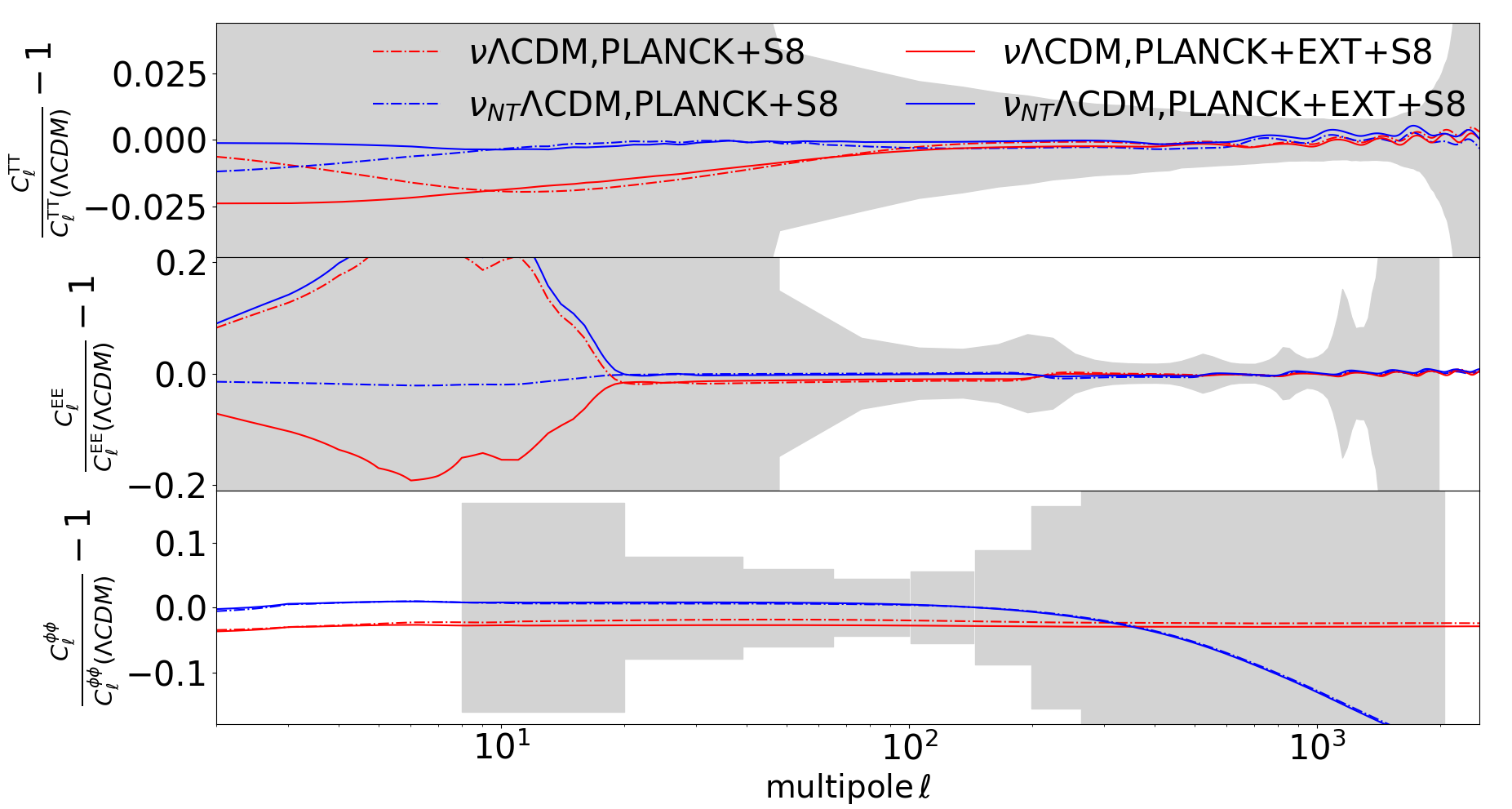}
\includegraphics[width=1.5\columnwidth]{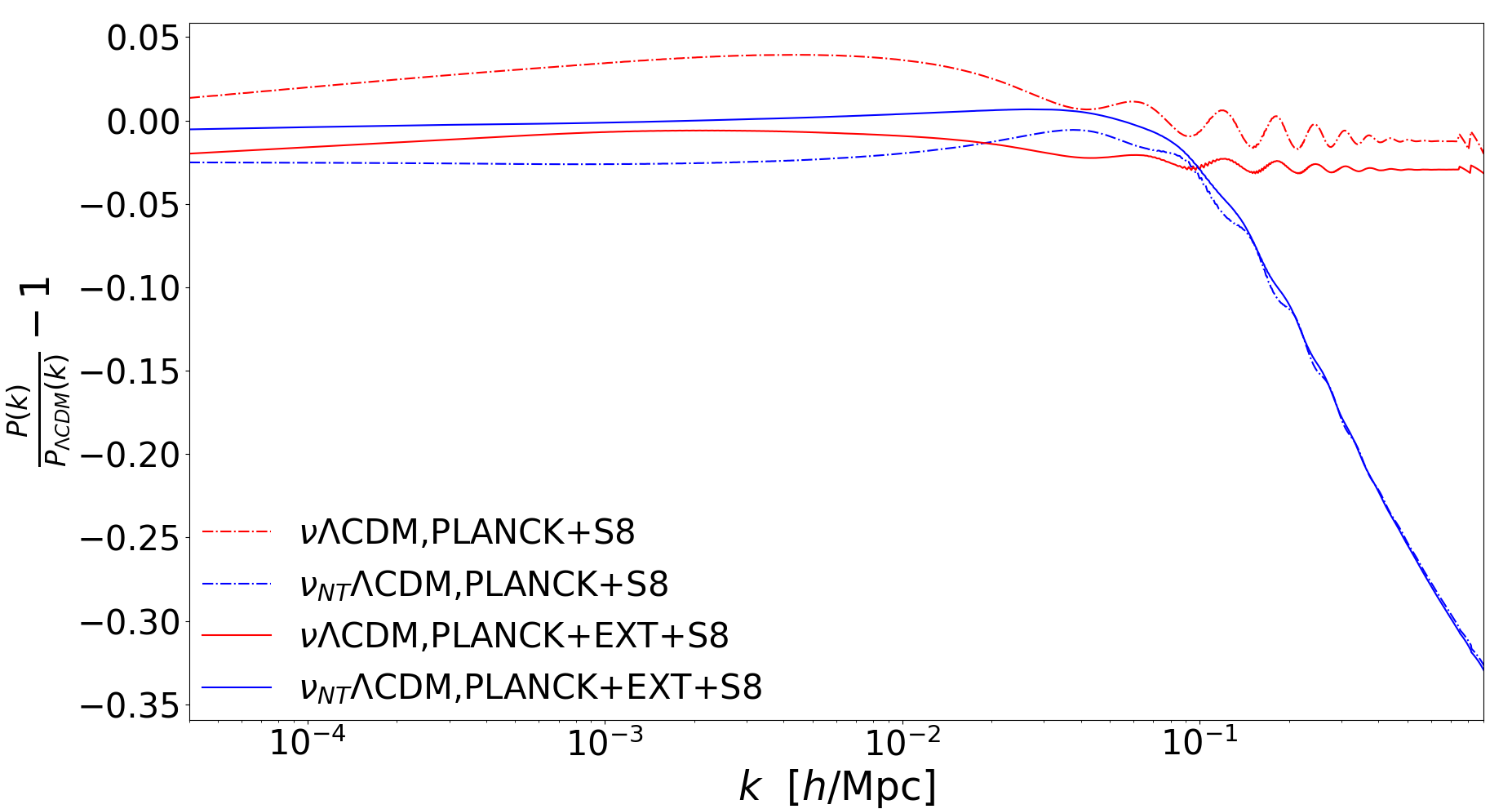}
\caption{Residuals of the CMB TT, EE, lensing (top panel) and matter (bottom panel) power spectra with respect to $\Lambda$CDM in the bestfit $\nu\Lambda$CDM and $\nu_{\rm NT}\Lambda$CDM models for two different datasets (see legend). The `Ext' data refers to BAO/FS+SN1a. \label{fig:cl_pk}}
\end{figure*}

\subsection{Understanding the MCMC}

\begin{figure*}[htp!]
\centering
\includegraphics[width=1.5\columnwidth]{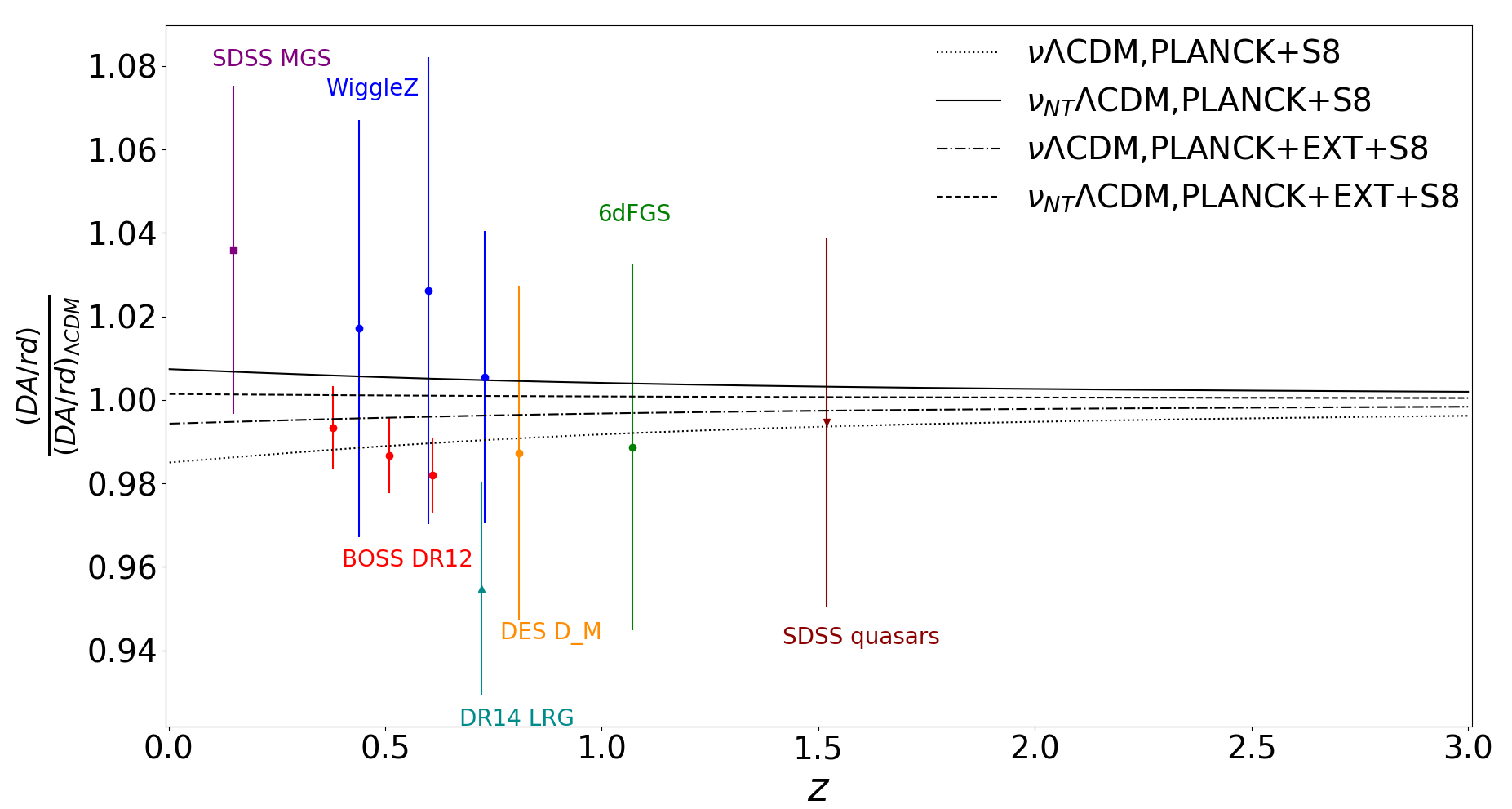}
\includegraphics[width=1.5\columnwidth]{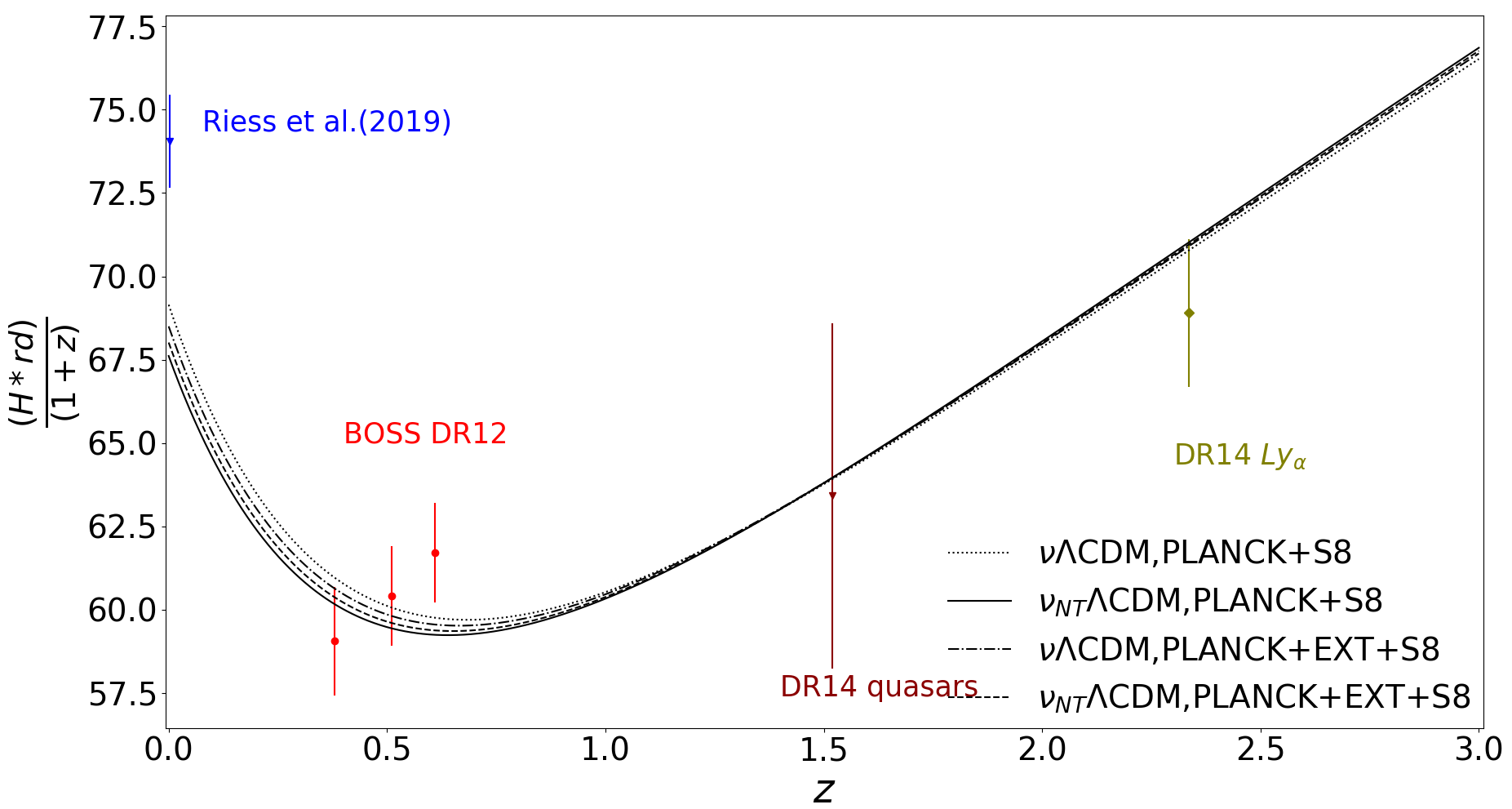}
\includegraphics[width=1.5\columnwidth]{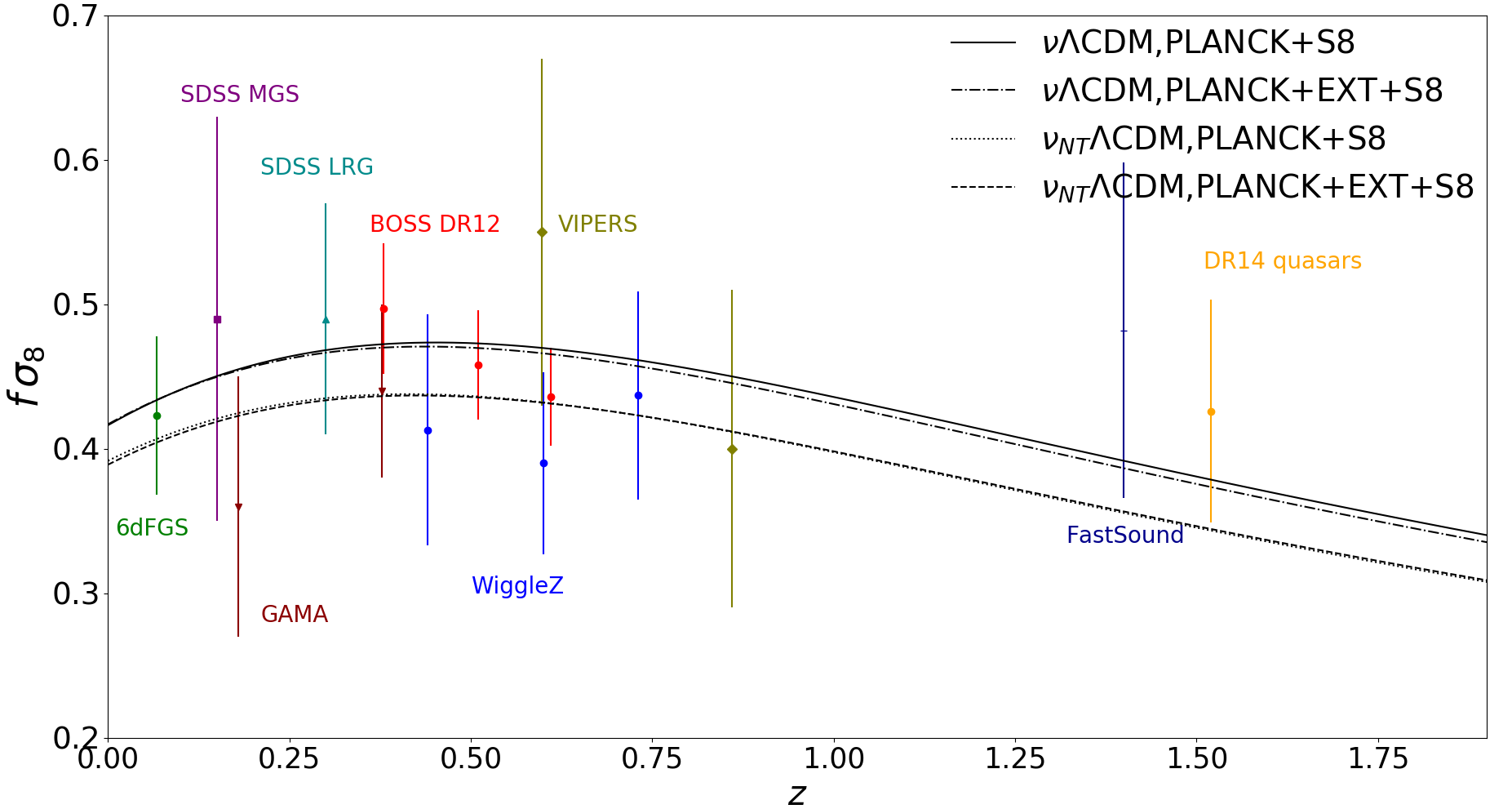}

\caption{Transverse BAO (top  panel), longitudinal BAO (middle panel) and growth factor (bottom panel) in the bestfit $\nu\Lambda$CDM and $\nu_{\rm NT}\Lambda$CDM models for two different datasets (see legend). The `Ext' data refers to BAO/FS+SN1a. The transverse BAO has been normalized to the $\Lambda$CDM prediction, as in Ref.~\cite{Akrami:2018vks}. \label{fig:bao_fs}}
\end{figure*}

 \begin{figure*}[htp!]
\centering

\includegraphics[width=1.4\columnwidth]{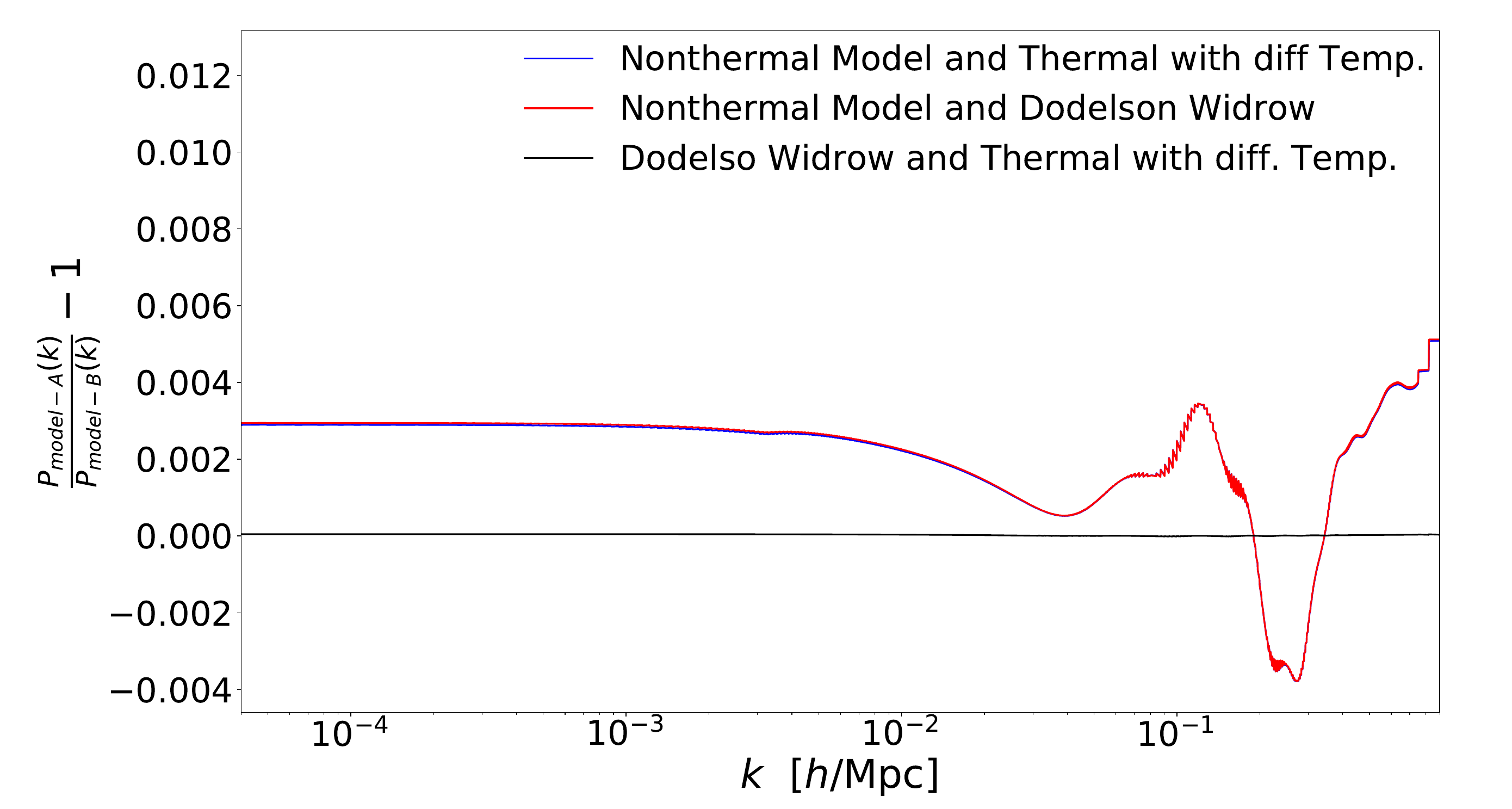}
\caption{Residuals of Matter power spectra  for various models (see legend). \label{fig:nt-vs-othermodels-pk}}
\end{figure*}

 \begin{figure*}[htp!]
\centering

\includegraphics[width=1.4\columnwidth]{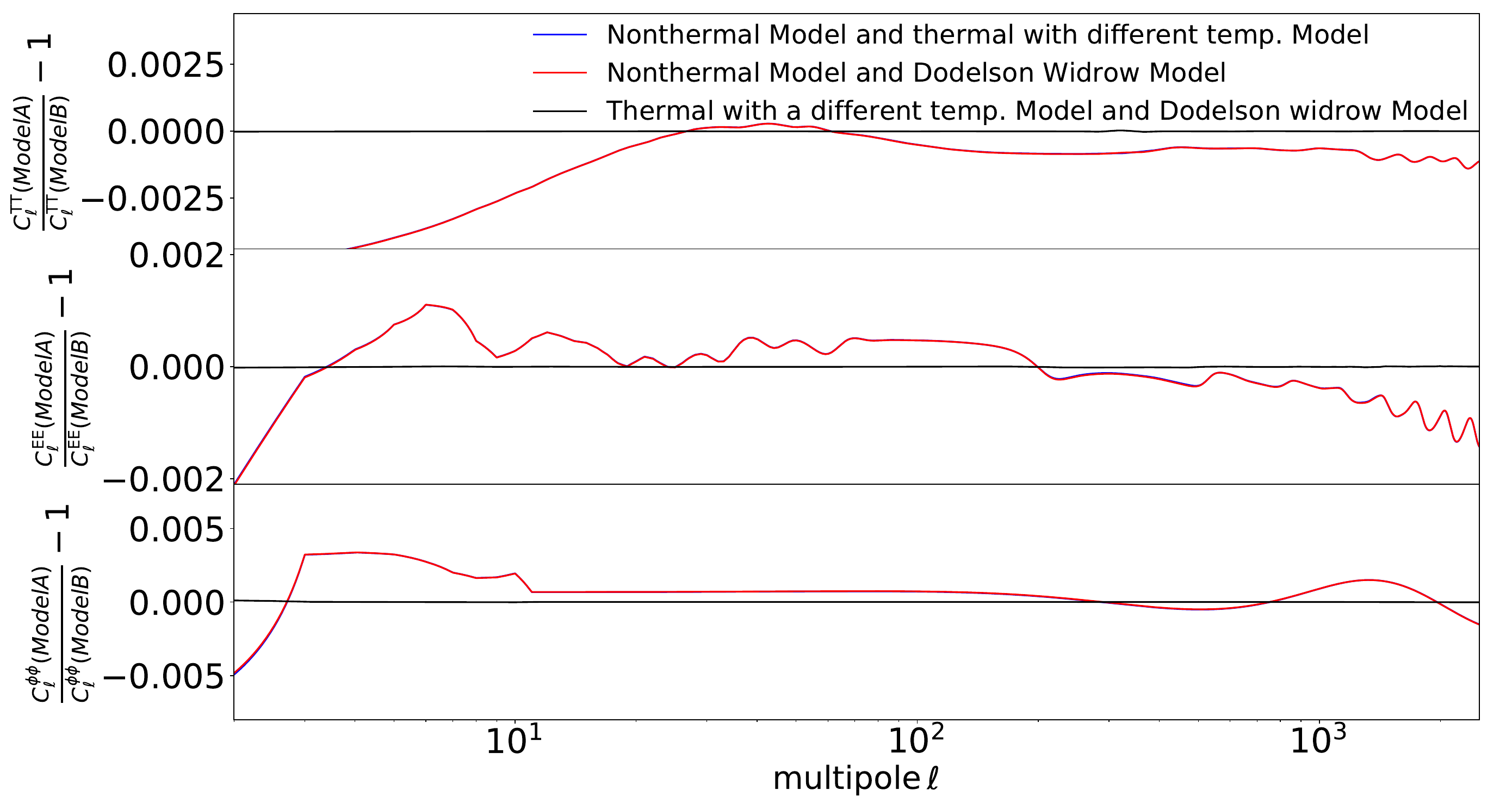}
\caption{Residuals of $C_l$ TT, TE and EE power spectra for various models (see legend). \label{fig:nt-vs-othermodels-cls}}
\end{figure*}

To understand better the results of the MCMC analyses, we show in Fig.~\ref{fig:cl_pk} the residuals of the CMB TT, EE, lensing (top panel) and matter (bottom panel) power spectra with respect to $\Lambda$CDM in the best fit $\nu\Lambda$CDM and $\nu_{\rm NT}\Lambda$CDM models obtained when considering Planck+S8 and Planck+Ext+S8 data.
We also show in Fig.~\ref{fig:bao_fs} the corresponding transverse BAO (top panel), longitudinal BAO (middle panel) and growth factor (bottom panel).
The first thing to notice is that, for a similar effect in the CMB power spectra, the corresponding power suppression in the matter power spectrum is much stronger in the $\nu_{\rm NT}\Lambda$CDM than in the $\nu\Lambda$CDM model. This is the reason why the $\nu_{\rm NT}\Lambda$CDM can perform much better in resolving the $S_8$ tension. 

Looking at the BAO and $f\sigma_8$ prediction, one can see that the most important difference is in the latter, which is significantly lower at all $z$ in the $\nu_{\rm NT}\Lambda$CDM because of this power suppression. This explains the small degradation in $\chi^2$ in the combined analysis with $S_8$. Moreover, the reconstructed dark matter density $\omega_{\rm cdm}$ in the $\nu_{\rm NT}\Lambda$CDM is also shifted by roughly $\sim1\sigma$ downwards (to compensate for the higher energy density due to the non-relativistic transition of the non-thermal neutrinos), which also leads to a small degradation in the fit to Planck data (hardly visible by eye in CMB power spectra residuals). 
This small difference in the matter density is also visible in the small$-k$ (large scales) branch of the matter power spectrum, particularly sensitive to $\Omega_m$ \cite{Lesgourgues:2018ncw}.
While these differences do not yet unambiguously rule out the $\nu_{\rm NT}\Lambda$CDM as a resolution to the $S_8$ tension, they do provide an interesting avenue to probe the model with future data, in particular through accurate measurements of the matter power spectrum, CMB lensing power spectrum and growth factor $f\sigma_8$.
An potential way to improve over the $\nu_{\rm NT}\Lambda$CDM results presented here is to assume that the hot component comes from the decay of a meta-stable cold dark matter species in the late-universe \cite{Abellan:2020pmw,Abellan:2021bpx}, instead of being present at all times. 
A good fit to all data can then be obtained when the mass-ratio of the mother and daughter particle $\varepsilon \sim 0.007$ and the CDM lifetime $\tau\sim~55$ Gyrs. 

\section{Implications for other non-thermal hot dark matter models}

\label{disc}

\begin{table*}[t]
    \centering
    \scalebox{0.9}{
    \begin{tabular}{|l|c|c|c|c|c|c|} 
    \hline
    Model & \multicolumn{2}{|c|}{Non-thermal} & 
\multicolumn{2}{|c|}{Thermal} & \multicolumn{2}{|c|}{Dodelson Widrow}  \\
\hline 
     \hline Data set &$ m_{\rm sp}$ [eV] & $B_{\rm sp}$ & $m_{\rm sp}$ [eV]& $\frac{T_s}{T_{\nu}}$ & $m_{\rm sp}$ [eV]& $\chi $\\         
     \hline 
    Planck  &  0.05 &  0.01&0  & 0.40  & 0& 0.03\\
    \hline 
     Planck+$S_8$  & 38.62&0.012 & 11.36  & 0.43  & 26.43& 0.03\\
    \hline
     Planck+Ext  &  18.98 &0.01 & 04.59& 0.36  & 12.85& 0.02\\
    \hline
     Planck+Ext+$S_8$  & 39.81 &0.01 & 11.75&0.43 &27.49& 0.03\\
    \hline
    \end{tabular}
    }
    \caption{Best-fit values of the physical parameters in the non-thermal, thermal and Dodelson-Widrow sterile neutrino models derived from our analyses. \label{tab:other_models}}
    
\end{table*}

  As discussed in the introduction and section \ref{macpi} , any distribution with the same values of $\nef$ and $\mef$ as ours should also relax the $\sigma_8$ tension. 
  Our results can thus be used to extract implications for the microscopic parameters of models which have momentum distributions different from the ones we have used.
  Here, we present such results for two models: 

a) Sterile particles at a different temperature from that
of the Standard Model neutrinos.
In this model, sterile neutrinos follow a thermal Fermi-Dirac Distribution.
\begin{equation}
    \label{FD_t}
    \hat{f}(p)=\frac{1 }{e^{p/T_{s}}+1}
\end{equation}
where $T_s$ is the temperature of sterile particles.
For a thermal sterile particle with a Fermi-Dirac distribution and a
different temperature $T_s$,
the quantities $\Delta N_{\rm eff}$  and $\omega_s$ become
\begin{equation}
    \label{DelN1}
    \Delta N_{\rm eff} = \left( \frac{T_{s}}{T_{\nu}^{\rm id}} \right)^4, \qquad
    \omega_s =\frac{m_{sp}}{94.05} \left(\frac{T_s}{T_\nu}\right)^{3}\,.
\end{equation}

b) The Dodelson-Widrow distribution \cite{Dodelson:1993je}
$$
\hat{f}(p) = {\chi \over {1 + e^{p/T_{\nu}}}}
$$
where  $T_{\nu}$ is the temperature of the neutrinos today, $\chi$ is a parameter related to the phenomenological parameters as \cite{acero}:
\begin{equation}
    \label{dw_1}
    \nef=\chi, \qquad
    \mef= m_{\rm sp} \times {\chi}\,,
\end{equation}
and $m_{\rm sp}$  is the individual neutrino mass in the model.

\newpage

\newpage

  We report the best-fit value of the model parameters in Tab.~\ref{tab:other_models}, obtained from translating our constraints on $\nef$ and $\mef$.
  We also show in Fig.~\ref{fig:nt-vs-othermodels-pk} and  \ref{fig:nt-vs-othermodels-cls}  the residuals of the matter power spectra and  CMB TT, TE, EE power spectra between our best-fit non-thermal HDM model and these two models. 
  This explicitly demonstrates our claim that, once $\nef$ and $\mef$ are fixed, observables are indistinguishable. 
  We note that the residuals between the thermal neutrino model at different temperatures and our non-thermal HDM model are of the order of the sensitivity of future LSS experiments such as EUCLID and LSST, and therefore this simple mapping might become limited in the future.  
  Note that, to avoid biasing constraints due to prior effects, we refrain from translating our reconstructed posterior on $\nef$ and $\mef$ into the model parameters.

The values we report in Tab.~\ref{tab:other_models} have  direct implication for thermalized hidden sector from both particle physics \cite{Reece:2015lch} and cosmological perspective \cite{Gariazzo:2013gua,Cyr-Racine:2013fsa}. Interestingly the main parameter for building a thermal hidden sector model is the temperature ratio 
$\xi= \frac{T_s}{T_{vis}}$ which received  a competitive constraint (though it depends on the model) from our analysis and  it may have strong implications for light sterile neutrino \cite{Gariazzo:2013gua} or other hidden sector particle physics models \cite{Foot:2014mia,Franca:2013zxa}. If the  hidden thermal particle interacts with dark matter or other particles in the dark sector, the coupling and other particle  physics parameters can be constrained from our result \cite{Brust:2013ova}.

It is tantalizing to connect the hot dark matter discussed here to the longstanding (and debated) short base line (SBL) anomalies \cite{Aguilar:2001ty,Aguilar-Arevalo:2013pmq} (see \cite{Maltoni,Dentler:2018sju} for recent reviews). Concretely, within the so called ``3+1'' neutrino scenario, those can be explained by a sterile neutrino with  $m_s \simeq\sqrt{\Delta m_{41}^2}~1$eV and a mixing angle leading to $\Delta N_{\rm eff}\simeq 1$. 
However, we find that the sterile particles required by the $S_8$-tension hints to a somewhat higher mass range $m_s\sim{\cal O}(10)$ eV (see tab.~\ref{tab:other_models}), and an almost negligible $\Delta N_{\rm eff}$. Our constraints, whether we include the $S_8$-prior or not, thus further confirm that a viable sterile neutrino solution to the SBL anomalies would require some additional mechanism to prohibit large $\Delta N_{\rm eff}$ production (see e.g.~\cite{Hamann:2011ge,Archidiacono:2016kkh,Chu:2015ipa,deSalas:2015glj} for examples). Nevertheless, it could be interesting to perform analysis including results from short baseline neutrino oscillation (e.g. with an additional prior as in Ref.~\cite{Gariazzo:2013gua}). This is beyond the scope of this paper and is kept for future study.

Finally, we also note that including data from the Bicep2/Kek array \cite{Array:2015xqh,Ade:2018gkx}, SPT-3G \cite{Balkenhol:2021eds} or ACT \cite{Aiola:2020azj} could help further constrain the sterile neutrino parameters thanks to higher accuracy measurement of the CMB damping tail and lensing spectrum. We also keep that for a future study, but refer to Refs.~\cite{Choudhury:2018sbz,Balkenhol:2021eds} for examples (constraints typically increases by $\sim10\%$, without considering a prior on $S_8$).


\section{Discussion and conclusions}
\label{sec:concl}
In this paper, we have explored the possibility that the `$S_8$-tension', the long-standing discrepancy between the  determination of the amplitude of the matter fluctuations from local \cite{Heymans:2013fya,MacCrann:2014wfa,
Hildebrandt:2018yau,Joudaki:2019pmv,
Abbott:2017wau,
Heymans:2020gsg} and cosmological \cite{Cosmo_collaboration2018planck} probes, is due to the existence of a non-thermal HDM contributing to a fraction of the DM density in the universe and leading to a power suppression at small-scales in the matter power spectrum.
Concretely, we have considered non-thermal HDM  produced as decay products of the inflaton. Such particles have the
momentum distribution  associated with decays
taking place in a matter-dominated universe evolving to radiation domination, as shown in  \cite{decayy}. However, we have argued that any model leading to the same $\nef$ and $\mef$ as our model (barring additional new physics ingredients) would lead to similar effects on cosmological observables, and therefore our constraints generically apply to any HDM models.

We have performed a comprehensive monte-carlo markov chains (MCMC) analysis against up-to-date data from Planck, BOSS (BAO and $f\sigma_8$) and Pantheon data, with and without the inclusion of a prior on the value of $S_8$ as measured with the KiDS/Viking+BOSS+2dFLens data.
Our findings can be summarized as follows:
\begin{enumerate}
\item[\textbullet] the $\nu_{\rm NT}\Lambda$CDM model can indeed alleviate the tension between Planck and $S_8$ measurements, but the success of the resolution is degraded once  BOSS and Pantheon data are included in the analysis.

\item[\textbullet] Compared to standard thermal neutrinos, the  $\nu_{\rm NT}\Lambda$CDM leads to a much stronger suppression in the matter power spectrum at late-times for a similar effect on the CMB power spectrum, and therefore to a more significant decrease in $\sigma_8$.

\item[\textbullet] The impact of the $\nu_{\rm NT}\Lambda$CDM is barely visible on the BAO scale and luminosity distance, but it does affect $f\sigma_8$ predictions. The model is, therefore  (somewhat) constrained by current BOSS growth factor measurements, and future measurements of the matter power-spectrum and $f\sigma_8$ at late-times will further test this scenario.

\item[\textbullet] We further discussed the connection between our model and generic phenomenological parameters constrained by the data that can be easily used to translate our constraints onto other similar models. Especially, we put constraints on other non-thermal HDM models--like the Dodelson-Widrow models  or on a thermal sterile particle with a different temperature in the hidden sector. We report competitive constraints on the
hidden sector temperature and DW scaling parameter which can have interesting particle physics implications, for instance in the context of SBL anomalies \cite{Aguilar:2001ty,Aguilar-Arevalo:2013pmq,Maltoni,Dentler:2018sju}.
\end{enumerate}

\begin{figure*}
    \centering
    \includegraphics[scale=0.35]{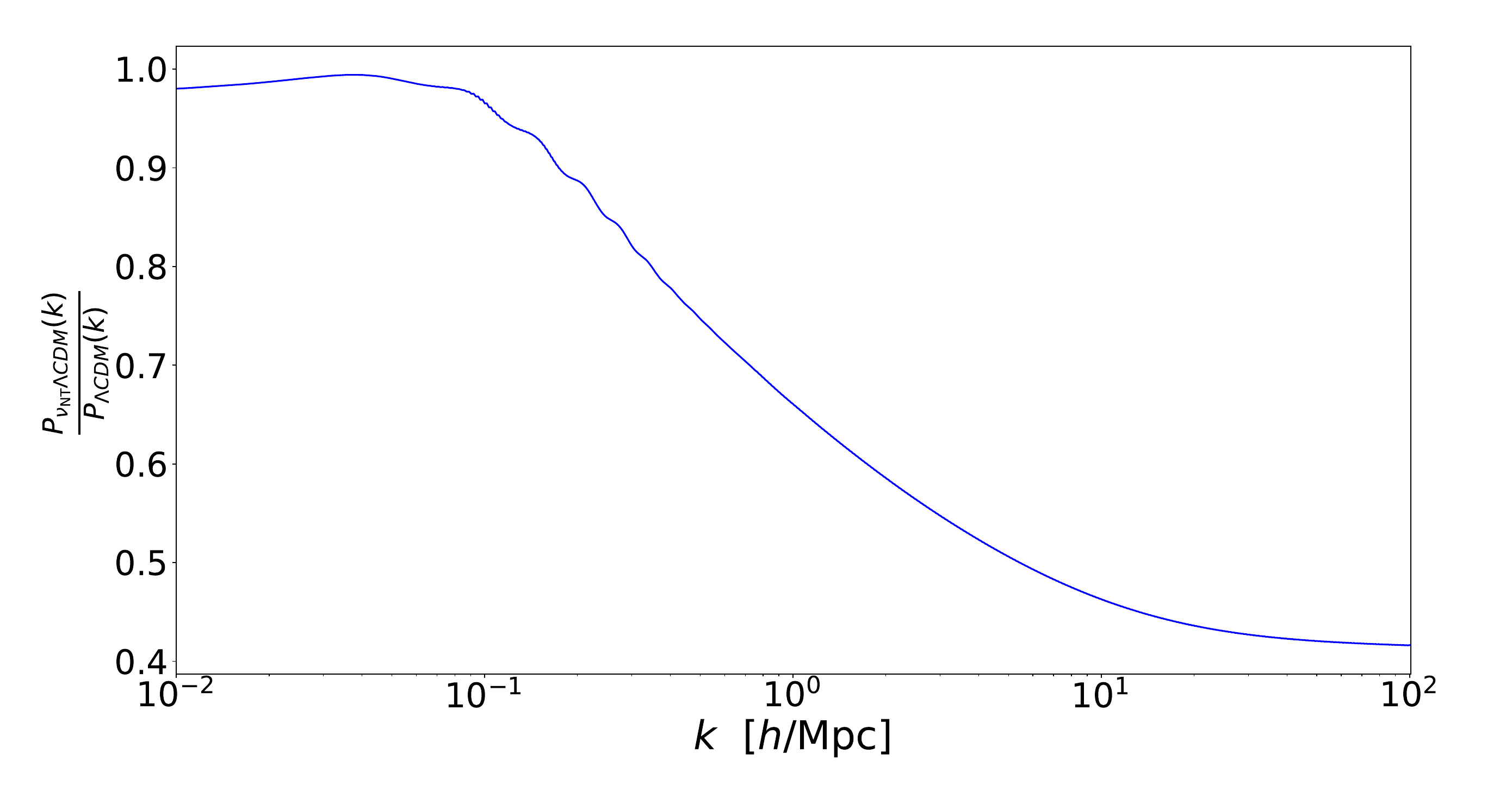}
    \caption{Ratio of matter power spectra in the $\nu_{\rm NT}\Lambda$CDM model to the $\Lambda$CDM model.}
    \label{fig:100_pk}
\end{figure*}

 It will be interesting to confront this model to Lyman-$\alpha$ forest flux power spectrum data along the lines of recent works \cite{Wang:2013rha,Murgia:2017lwo,Baur:2017stq,Murgia:2018now,Archidiacono:2019wdp,Miller:2019pss,Palanque-Delabrouille:2019iyz,Enzi:2020ieg}. For instance, Ref.~\cite{Murgia:2018now} established that any non-cold DM scenario must leave the spectrum at $k\lesssim 33h/$Mpc unaffected. The model studied, whose spectrum shows deviation already at $k\sim 0.05-1h/$Mpc, could therefore likely be probed by Lyman-$\alpha$ data.  Nevertheless, the non-thermal neutrino only represents a small fraction of the total DM density, and constraints do not necessarily trivially apply on the model, since the suppression stops at large $k$'s. This is explicitly shown in the Fig.~\ref{fig:100_pk}, where we compare the linear prediction of the matter power spectrum for the $\Lambda$CDM and $\nu_{\rm NT}\Lambda$CDM model at scales up to $k=100h/$Mpc. For instance, Ref.~\cite{Boyarsky:2008xj} derived constraints on WDM+CDM models, showing that model with similar level of suppression at high$-k$'s than the one studied here are allowed by the data (although an analysis of more recent data is still lacking). Additionally, the latest study dealing with thermal warm dark matter and neutrinos, has established a mild-tension ($3\sigma$) between lyman-$\alpha$  and Planck data \cite{Palanque-Delabrouille:2019iyz}. {\color{red}} In the context of the $\sigma_8$ tension, it would therefore be interesting to check in detail whether a non-thermal hot dark matter model can play a role in alleviating the ``lyman-$\alpha$ tension'' \footnote{We note that approximate bounds could be computed using a formalism relating the constraints on effective parameters between models (see e.g. Ref.~\cite{Ballesteros:2020adh}). However, this would be missing the possibility that the model helps in resolving the tension, and therefore it is worth looking into it in more details than this matching.}. 
 Additionally, future high accuracy measurement of the matter power spectrum at small scales by upcoming surveys such as  Euclid \cite{Amendola:2016saw}, LSST \cite{Mandelbaum:2018ouv}, and DESI \cite{Aghamousa:2016zmz} can further test these models as a resolution to the $S_8$-tension.

\section*{acknowledgements}

We thank Guillermo Franco Abell\`an and Riccardo Murgia for many interesting discussions. {\color{red} We thank the anonymous referee for useful comments that help improve our paper.}
AM is supported
in part by the SERB, DST, Government of India by the grant MTR/2019/000267. SD acknowledges SERB DST Government of India grant  CRG/2019/006147 for supporting the project.
VP is partly supported by the CNRS-IN2P3 grant Dark21  and by the European Union’s Horizon 2020 research and innovation program under the Marie Skodowska-Curie grant agreement No 860881-HIDDeN.  The authors acknowledge the use of computational resources from the Dark Energy computing Center funded by the OCEVU Labex (ANR-11-LABX-0060) and the Excellence Initiative of Aix-Marseille University (A*MIDEX) of the “Investissements d’Avenir” programme as well as IIA Nova cluster where initial analysis was carried out. 

\appendix

\section{$\chi^2_{\rm min}$ per experiment}
\label{app:chi2}
We report $\chi^2_{\rm min}$ per experiment in each of the analysis performed.

\begin{table}[hbt!]
\scalebox{0.9}{
  \begin{tabular}{|l|c|c|c|c|}
  \hline
Experiment & 
  \multicolumn{4}{|c|}{$\Lambda$CDM} \\
  \hline
  Planck~high$-\ell$ TT,TE,EE &2346.7 &  2350.8 &  2346&2349.1   \\
Planck~ low$-\ell$ EE & 396 &396.1 & 396.8 &  396.2  \\
Planck~ low$-\ell$ TT  &23.2 & 22.5 & 23.4 &  22.6  \\
Planck~lensing  &8.8 & 9.6& 9.2  &   9.1 \\
Pantheon  &$-$ & $-$ &  1026.9 &   1026.7\\
BAO/FS~BOSS DR12  &$-$ &$-$ &  6.9 &   6.5\\
BAO~BOSS low$-z$  & $-$& $-$ & 1.2 &  2.3 \\
KiDS/BOSS/2dFGS & $-$& 4.6&  $-$ &  5.9  \\

  \hline
total &2774.8 &2783.4& 3810.4&  3818.2\\
  \hline

  \end{tabular} 
  }
  \caption{Best-fit $\chi^2$ per experiment (and total) in the $\Lambda$CDM model.}
    \label{tab:chi2_lcdm}

\end{table}

\begin{table}[hbt!]
\scalebox{0.9}{
  \begin{tabular}{|l|c|c|c|c|}
  \hline
Experiment & 
  \multicolumn{4}{|c|}{$\nu\Lambda$CDM} \\
  \hline
  Planck~high$-\ell$ TT,TE,EE & 2345.98&2348.2  & 2346.9 &2348.6  \\
Planck~ low$-\ell$ EE &  396.54&396.8  &396.5  &  395.7 \\
Planck~ low$-\ell$ TT  &23.3 &22.2  & 22.8 & 22.4    \\
Planck~lensing  &9.03 &  8.9&  8.8& 9.3  \\
Pantheon  &$-$ &  $-$ & 1026.8 &  1026.7 \\
BAO/FS~BOSS DR12  &$-$ &  $-$ & 6.1 &  5.9 \\
BAO~BOSS low$-z$  & $-$&$-$   &  1.7&  1.7 \\
KiDS/BOSS/2dFGS & $-$&  5.8& $-$ &  6.1  \\

  \hline
total &2774.9 &2782.0 & 3809.5& 3816.4 \\
  \hline

  \end{tabular} 
  }
  \caption{Best-fit $\chi^2$ per experiment (and total) in the model with massive thermal neutrinos and additional relativistic degrees of freedom.}
    \label{tab:chi2_nu}

\end{table}

\begin{table}[hbt!]
\scalebox{0.9}{
  \begin{tabular}{|l|c|c|c|c|}
  \hline
Experiment & 
  \multicolumn{4}{|c|}{$\nu_{\rm NT}\Lambda$CDM} \\
  \hline
  Planck~high$-\ell$ TT,TE,EE &  2346.7 & 2348.7&2 2346.4&2349.1\\
Planck~ low$-\ell$ EE & 396.3 &  395.9&396.8 &396.9   \\
Planck~ low$-\ell$ TT  & 23.1 & 23.3&23.4 &  23.1 \\
Planck~lensing  & 8.8 &  9.2&  8.8&   9.1 \\
Pantheon  & $-$ &$-$ &1026.8&   1026.7\\
BAO/FS~BOSS DR12  &  $-$& $-$ & 6.1&    6.8\\
BAO~BOSS low$-z$  &  $-$&$-$& 1.4 & 1.7 \\
KiDS/BOSS/2dFGS &  $-$&1.6 &$-$ &    1.2 \\

  \hline
total & 2775.0 &2778.6 &  3809.7 &  3814.5\\
  \hline

  \end{tabular} 
  }
  \caption{Best-fit $\chi^2$ per experiment (and total) in the non-thermal sterile neutrino model.}
    \label{tab:chi2_nt}

\end{table}

\section{On the relationship between observables and effective parameters}
\label{modelY}

 The fact that the two parameters $\Delta N_{\rm eff}$ and $m_{\rm eff}$ determine the physical observables is well known, as mentioned already \cite{acero}. In fact, this  is
also used by the Planck collaboration for their analysis, see e.g figure 37, section 7.5.2 of \cite{Planck:2018vyg}. For completeness, in this appendix
we analyse this expectation in our setting.  In the model discussed in the main text  $m_{\varphi} = 10^{-6} M_{\rm pl}$ and $\tau = 10^{8}/ m_{\varphi}$
(we will refer to this as model X). Here, we consider $m_{\varphi} = 10^{-8} M_{\rm pl}$ and $\tau = 10^{9} \big{/} m_{\varphi}$ (we will refer to this as model Y).
 
\begin{figure*}[htp!]
\centering

\includegraphics[width=1.6\columnwidth]{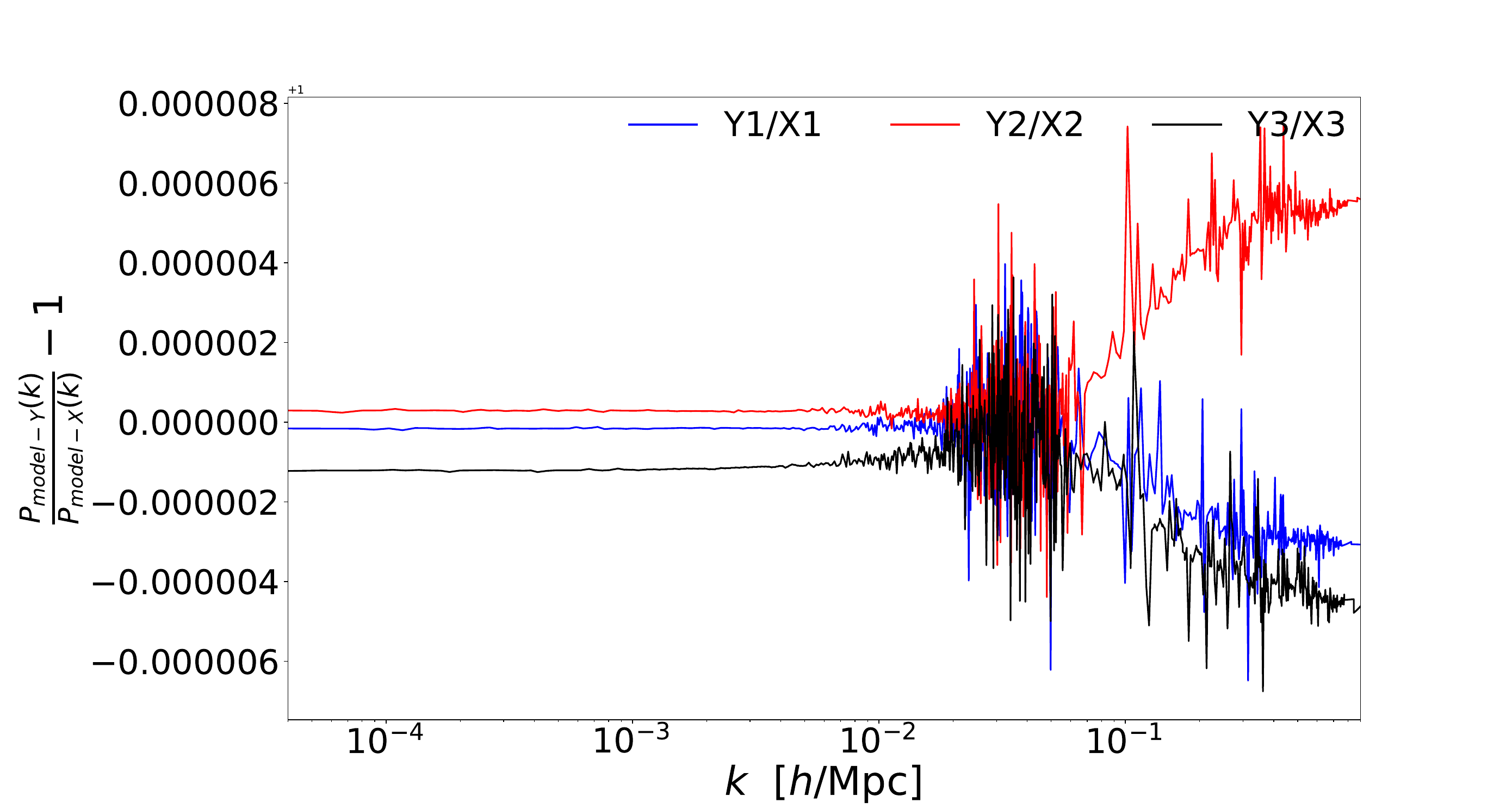}
\includegraphics[width=1.5\columnwidth]{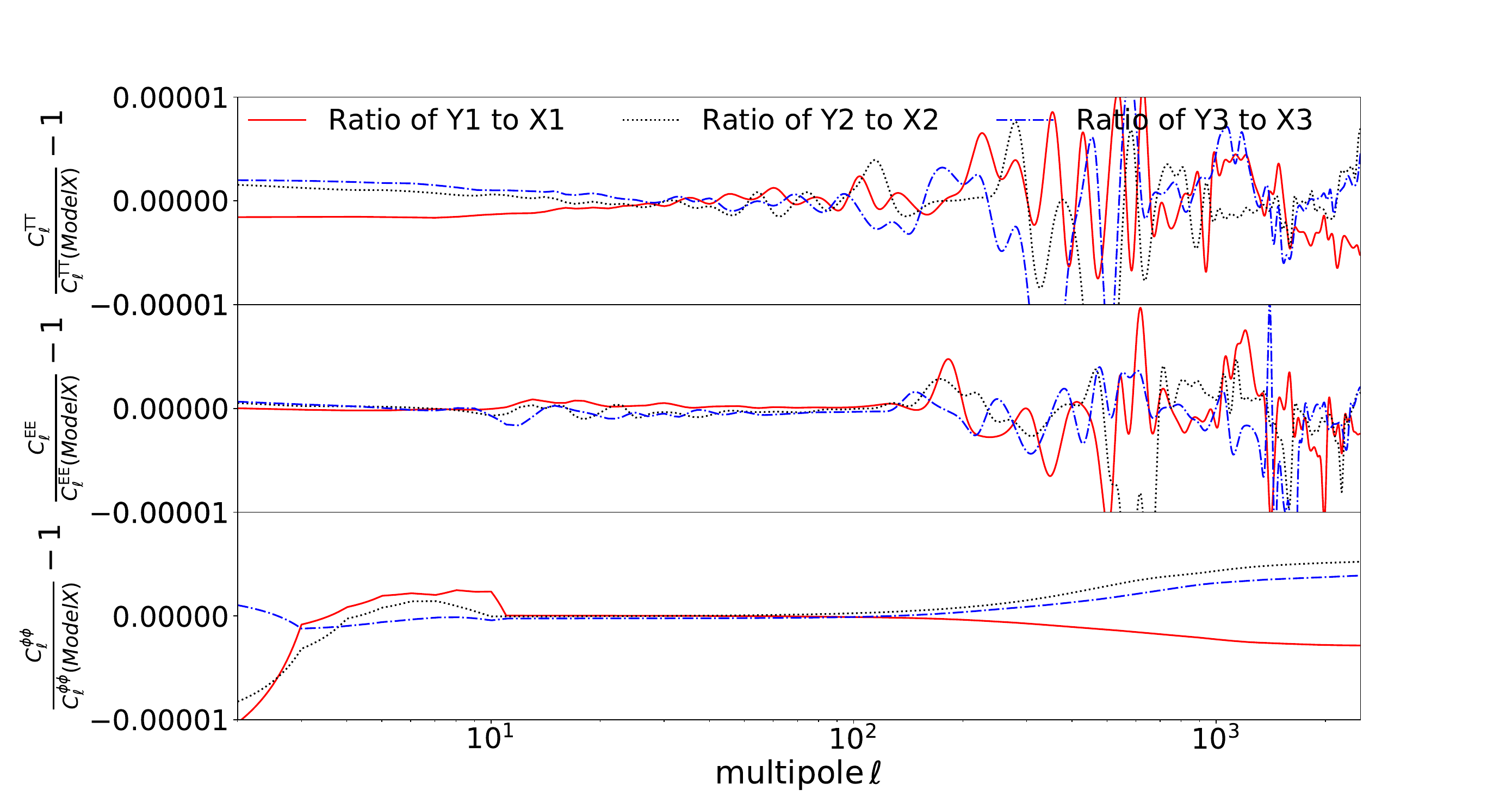}
\caption{Residuals of Matter power spectra and CMB TT EE $\phi \phi$ power spectra for various models (see legend). Here the models 
(X1,Y1),(X2,Y2),(X3,Y3) correspond to the models described in table \ref{tab:same_neff_tab}.
\label{fig:same_neff_1}} 
\end{figure*}

Note that equations \pref{delN} and \pref{meff} imply
that if $B^{Y}_{\rm sp} = B^{X}_{\rm sp}$ and $m^{Y}_{\rm sp} = m^{X}_{\rm sp} \big{/} \sqrt{10}$, models X and Y will have equal values of $\Delta N_{\rm eff}$ and $m_{\rm eff}$. 
We compare the CMB and matter power spectra today for equal values of $m_{sp}$ and $B_{sp}$  in figure \ref{fig:same_neff_1}.
As expected, we find  that the CLASS inputs of models X and Y are in very good agreement (better than $10^{-5}$). Therefore, our constraints are robust to the specific choice of these parameters.

\begin{table*}[htp!]
    \centering
\begin{tabular}{ |c|c|c|c|c|c|c| } 

 \hline
  parameter & model X1 & model Y1 & model X2 & model Y2  & model X3 & model Y3 \\ 
 \hline
 $m_{\varphi} $& $10^{-6} M_{\rm pl}$ & $10^{-8} M_{\rm pl}$ & $10^{-6} M_{\rm pl}$ & $10^{-8} M_{\rm pl}$& $10^{-6} M_{\rm pl}$ & $10^{-8} M_{\rm pl}$ \\ 
 \hline
 $\tau$ & $10^{8} \big{/} m_{\varphi}$ & $10^{9} \big{/} M_{\rm \varphi}$  & $10^{8} \big{/} m_{\varphi}$ & $10^{9} \big{/} m_{\varphi}$ & $10^{8} \big{/} m_{\varphi}$ & $10^{9} \big{/} m_{\rm \varphi}$ \\ 
 \hline
 $m_{\rm sp}$ (in eV)  & 38.62194 & ${38.62194  \over \sqrt{10}}$ & 38.62194 & ${38.62194  \over \sqrt{10}}$ & 28.62194 & ${28.62194  \over \sqrt{10}}$ \\
 \hline
 $B_{\rm sp}$ & 0.0118 & 0.0118  & 0.0218 & 0.0218  & 0.0218 & 0.0218 \\
 \hline
\end{tabular}

    \caption{Table shows the parameters of three pairs of model (X1,Y1),(X2,Y2),(X3,Y3). Both the models of the each pair have different values of $\tau$ and $m_{\phi}$.}
    \label{tab:same_neff_tab}
\end{table*}

\pagebreak

\newpage

\bibliography{biblio}
\end{document}